%% file: charm2015_APalanoLHCb.tex
\documentclass[12pt]{article}
\usepackage{graphicx}
\usepackage{xspace}
\usepackage[percent]{overpic}
\usepackage{colordvi}
\usepackage{xcolor}

\newcommand\pubnumber{}
\newcommand\pubdate{\today}

\def\palano{INFN and University of Bari, Italy}
\def\support{\footnote{on behalf of the LHCb Collaboration.}}
\def\Title#1{\begin{center} {\Large #1 } \end{center}}
\def\Author#1{\begin{center}{ \sc #1} \end{center}}
\def\Address#1{\begin{center}{ \it #1} \end{center}}

\newcommand\pubblock{\rightline{\begin{tabular}{l} \pubnumber\\
         \pubdate  \end{tabular}}}
\newenvironment{Abstract}{\begin{quotation}  }{\end{quotation}}
\newenvironment{Presented}{\begin{quotation} \begin{center} 
             PRESENTED AT\end{center}\bigskip 
      \begin{center}\begin{large}}{\end{large}\end{center} \end{quotation}}

\input econfmacros.tex

\def\mev     {~MeV}
\newcommand{\gevc}{\ensuremath{{\mathrm{\,Ge\kern -0.1em V\!/}c}}\xspace}
\newcommand{\mevc}{\ensuremath{{\mathrm{\,Me\kern -0.1em V\!/}c}}\xspace}
\newcommand{\gevcc}{\ensuremath{{\mathrm{\,Ge\kern -0.1em V\!/}c^2}}\xspace}
\newcommand{\mevcc}{\ensuremath{{\mathrm{\,Me\kern -0.1em V\!/}c^2}}\xspace}

\def\Km      {\ensuremath{K^-}\xspace}
\def\pip      {\ensuremath{\pi^+}\xspace}

\def\PK      {\ensuremath{\mathrm{K}}\xspace}
  \def\Kbar    {{\kern 0.2em\overline{\kern -0.2em \PK}{}}\xspace}
\def\Kstarb  {{\ensuremath{\Kbar{}^*}}\xspace}

\def\Kp    {\ensuremath{K^+}\xspace}
\def\Km    {\ensuremath{K^-}\xspace}

\def\pip   {\ensuremath{\pi^+}\xspace}
\def\pim   {\ensuremath{\pi^-}\xspace}

\def\Dzb     {{\ensuremath{\bar{\D^0}}}\xspace}

\def\KS    {\ensuremath{K^0_{\scriptscriptstyle S}}\xspace}
\def\D       {\ensuremath{D}\xspace}

\def\DTwentyFiveFiftyNeutral {\ensuremath{D_{J}(2580)^0}\xspace}

\def\DTwentySevenFiftyNeutral {\ensuremath{D_{J}(2740)^0}\xspace}

\def\DThreeU {\ensuremath{D_{J}(3000)^0}\xspace}

\def\PK      {\ensuremath{\mathrm{K}}\xspace}
\def\kaon    {{\ensuremath{\PK}}\xspace}
\def\Kstar   {{\ensuremath{\kaon^*}}\xspace}

\newcommand{\al}{\ensuremath{\kern 0.5em }}
\newcommand{\all}{\ensuremath{\kern 0.25em }}

\begin{document}
\begin{titlepage}
\pubblock

\vfill
\Title{Open charm Spectroscopy and exotic states at LHCb}
\vfill
\Author{Antimo Palano\support}
\Address{\palano}
\vfill
\begin{Abstract}
We present a summary of new experimental results on the status of the charm spectroscopy using inclusive approaches and Dalitz plot analyses of $B$ and $B_s$ decays. We also report on a new determination of the $X(3872)$ quantum numbers.
\end{Abstract}
\vfill
\begin{Presented}
The 7th International Workshop on Charm Physics (CHARM 2015)\\
Detroit, MI, 18-22 May, 2015
\end{Presented}
\vfill
\end{titlepage}
\def\thefootnote{\fnsymbol{on behalf of the LHCb Collaboration}}
\setcounter{footnote}{0}
%

\section{Introduction: Charm meson spectroscopy}
The quark model predicts many states with different quantum numbers in limited mass regions~\cite{isgur1,isgur2}.
New progress in the understanding of the $D_J$ and $D_{sJ}$ spectra in LHCb experiment come from:
    \begin{itemize}
      \item{} Inclusive studies: study of the reactions $ p p \to D_J/D_{sJ} X$;
    \item{} Exclusive studies in Dalitz plot analyses of $B$ and $B_s$ decays.
    \end{itemize}
    In the following, we remind that states having $J^P=0^+,1^-,2^+,3^-, ..$ are defined as having ``Natural Parity'', while states having $J^P=0^-,1^+,2^-, ...$ are defined as having ``Unnatural Parity''.
A resonance decaying to $D \pi$ has ``Natural Parity''. Labeled with $D^*$.
The $D^*(\pi/K)$ system can access to both ``Natural Parity'' and ``Unnatural Parity'' states, except for $J^P=0^+$ which is forbidden.

In the following, inclusive studies make use of 1 $fb^{-1}$ while Dalitz analyses of 3 $fb^{-1}$ integrated luminosities.
\section{Results on $D_J$ mesons spectroscopy}

\subsection{Inclusive studies}

   \begin{figure}[htb]
\centering
\includegraphics[height=2.0in]{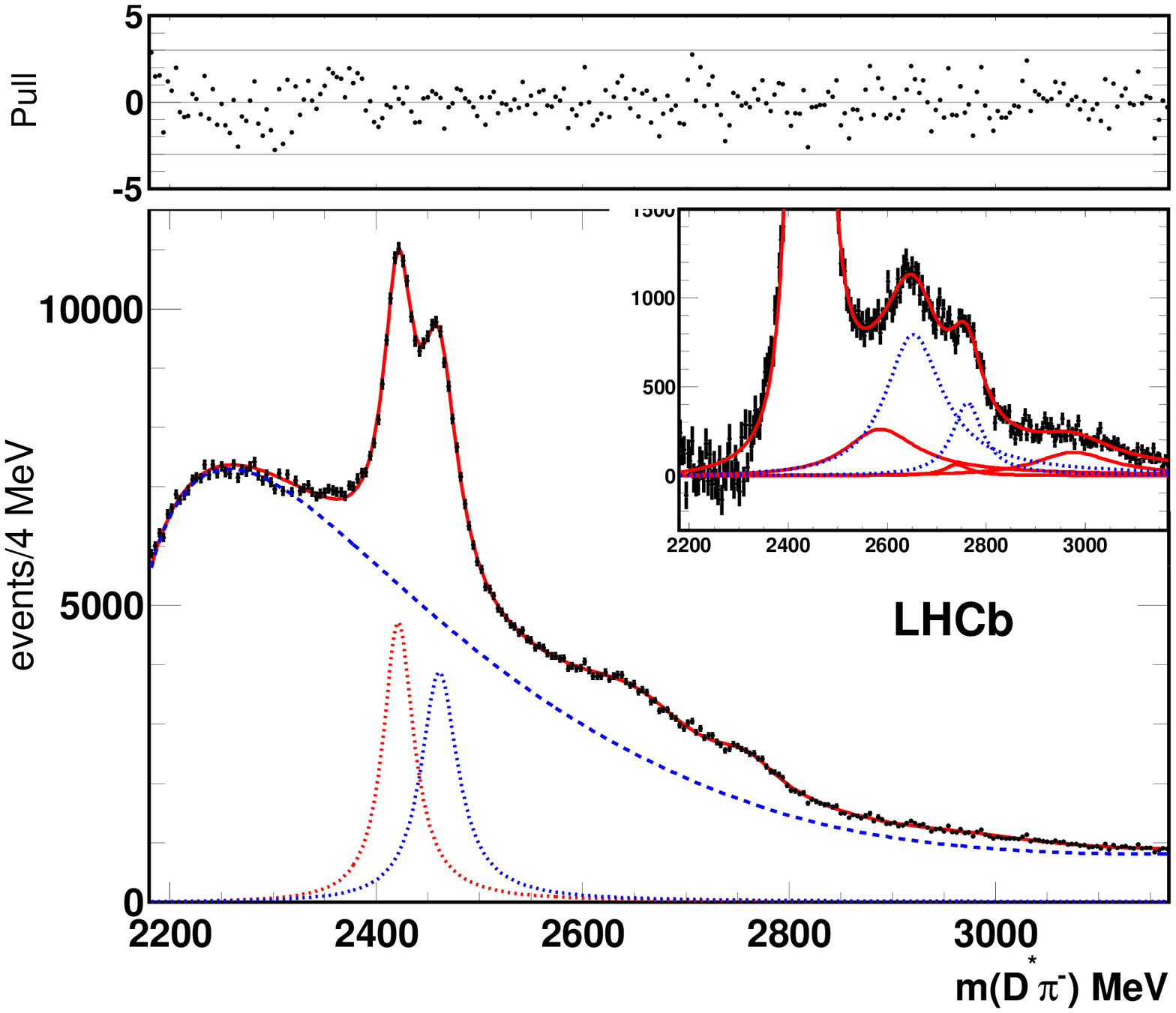}
\includegraphics[height=2.0in]{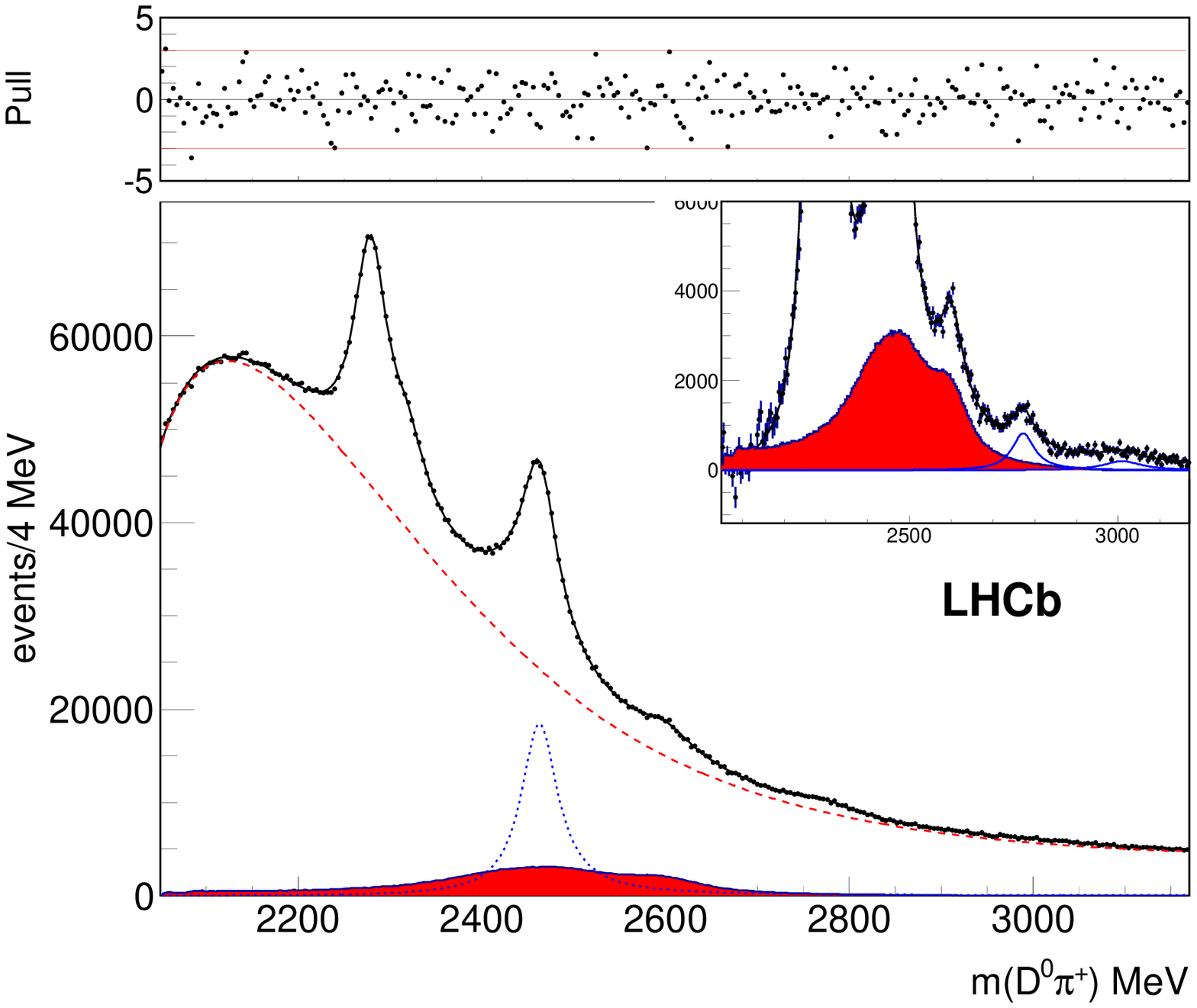}
\caption{Left: $D^{*+} \pim$ mass spectrum with enhanced natural parity selections. Right: $D^0 \pi^+$ mass spectrum.
Note the crossfeed (in red) from the high mass $D_J$ resonances into the $D \pi$ mass spectra.}
\label{fig:fig1}
   \end{figure}
   
The BaBar~\cite{djbabar} and LHCb~\cite{djlhcb} experiments observe two new natural parity resonances, $D^*(2650)$ and $D^*(2760)$, both decaying to $D \pi$ and $D^* \pi$. While the parameters of the $D^*(2760)$ are consistent within the two experiments, the mass  of  the $D^*(2650)$ state is shifted down by $\approx$ 40 MeV in the BaBar analysis due to
different handling of the $D^* \pi$ feedthrough into the $D \pi$ final states. The two states are candidates for being the $J^P=1^-$ $D_1^3(2S)$ and $J^P=1^-$ $D_1^3(1D)$. 

\begin{figure}[htb]
   \centering
   \includegraphics[width=3.0in]{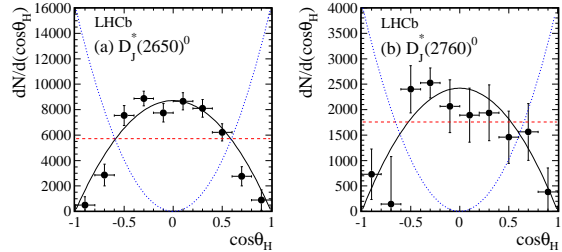}
   \caption{Angular distributions in terms of the helicity angle for the $D^*_J(2650)$ and $D^*(2760)$.}
   \label{fig:fig2}
  \end{figure}

\begin{figure}[htb]
\centering
  \includegraphics[width=3.5in]{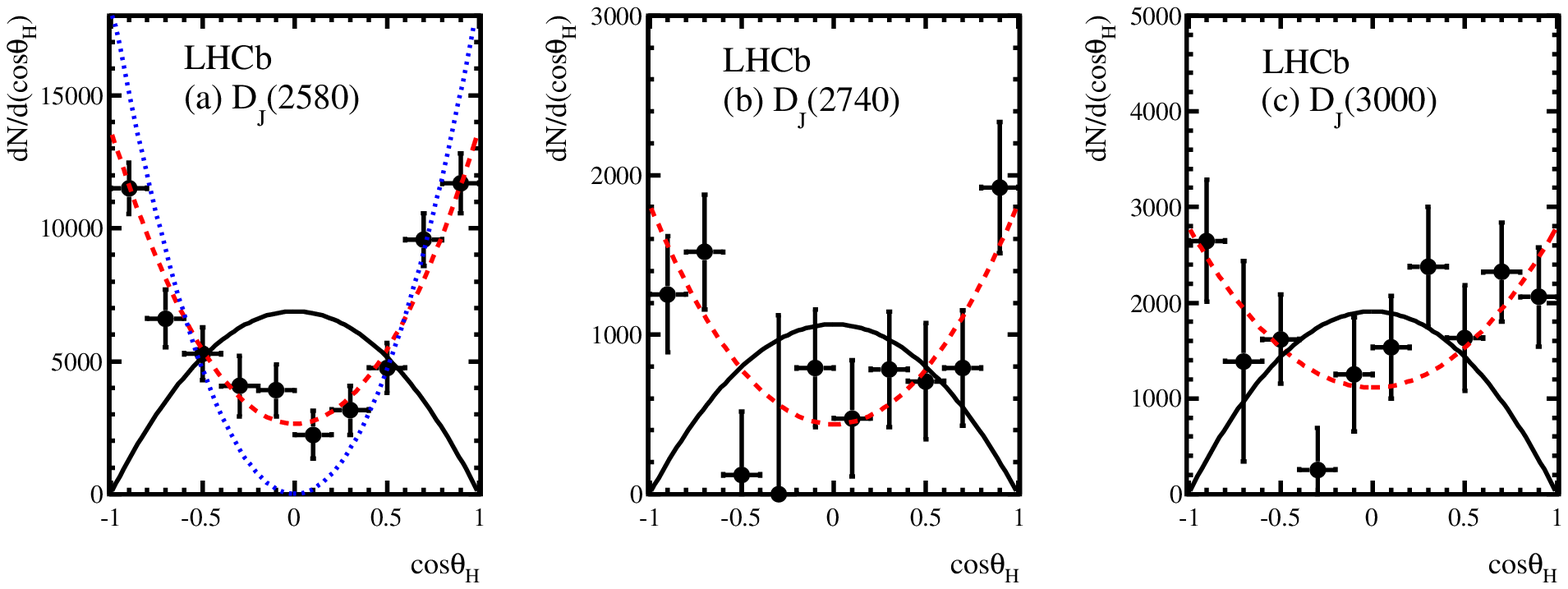}
   \caption{Angular distributions in terms of the helicity angle for the \DTwentyFiveFiftyNeutral, \DTwentySevenFiftyNeutral, and \DThreeU.}
   \label{fig:fig3}
 \end{figure}

 Adding statistical and systematic uncertainties in quadrature, we obtain a weighted mean values for $D^*_J(2770)$ parameters.
 \begin{center}
   $m(D^*_J(2770))=2768.8 \pm 1.7 \ MeV, \qquad \Gamma(D^*_J(2770)) = 63.2 \pm 5.3 \ MeV$
 \end{center}
 The $D^{*+} \pim$ angular distributions in terms of the helicity angle for the $D^*_J(2650)$ and $D^*(2760)$ are shown in fig.~\ref{fig:fig2}. They are well fitted by the $sin^2\theta_H$ functions and therefore are consistent with having
 natural parity.
LHCb experiment also observes three unnatural parity states, \DTwentyFiveFiftyNeutral, \DTwentySevenFiftyNeutral, and \DThreeU, whose angular distributions are shown in fig.~\ref{fig:fig3}.
 The weighted mean values for $D_J(2580)$ parameters, consistent with a $J^P=0^-$ assignment, are
 \begin{center}
   $m(D_J(2580))=2564.0 \pm 5.1 \ MeV, \qquad \Gamma(D_J(2580)) = 135.6 \pm 16.9 \ MeV$
 \end{center}
 The weighted mean values for the $D_J(2740)$ parameters are
  \begin{center}
    $m(D_J(2740))=2751.3 \pm 3.1 \ MeV, \qquad \Gamma(D_J(2740)) = 71.4 \pm 11.4 \ MeV$
  \end{center}  
  and is a candidate for being a $J^P=2^-$ state. The broad structures observed in the 3000 MeV mass region could be
  a superposition of several states.

  \subsection{First observation and Dalitz plot analysis of $B^- \to D^+ K^- \pi^-$}
     
\begin{figure}[htb]
\centering
  \includegraphics[width=2.5in]{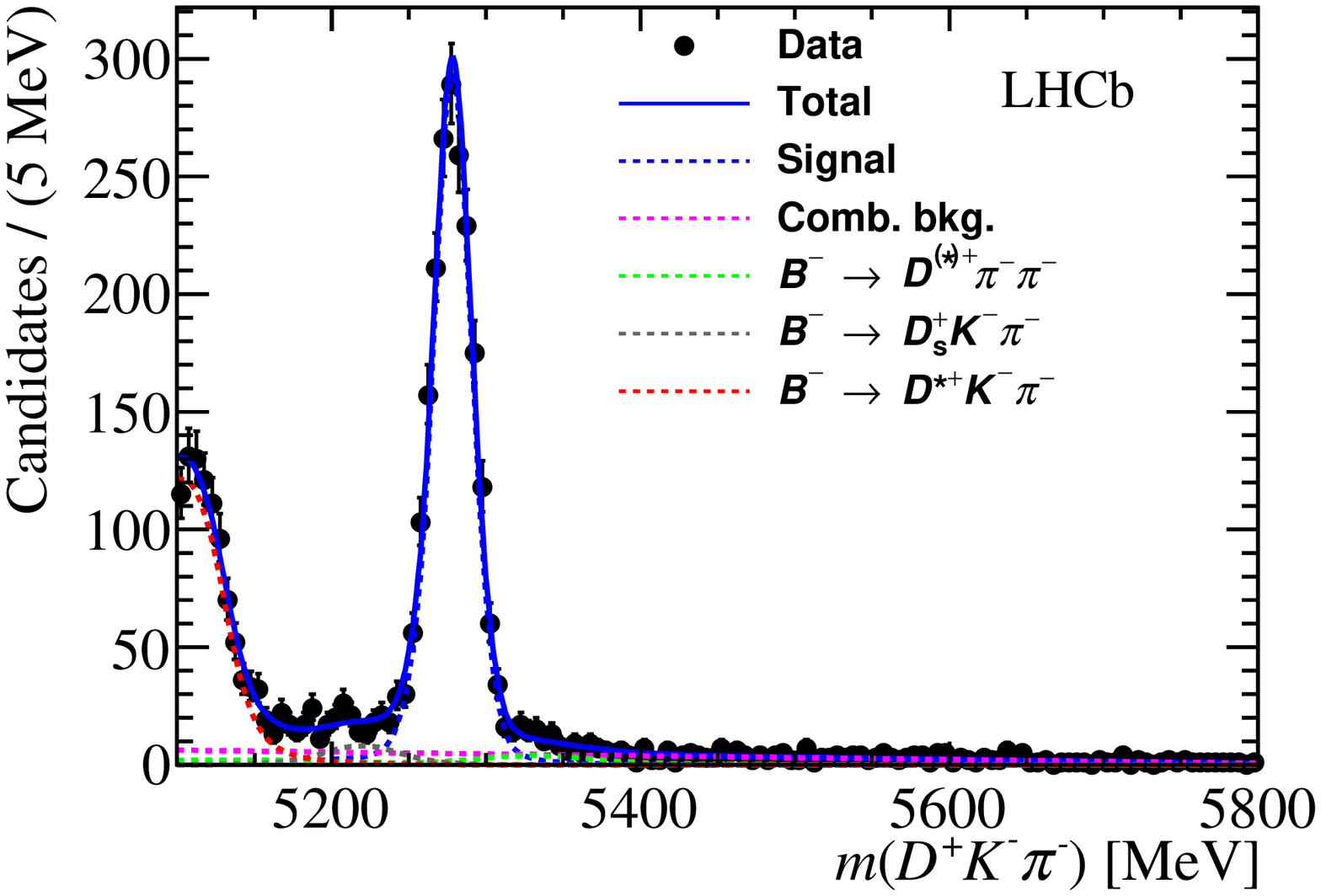}
  \includegraphics[width=2.5in]{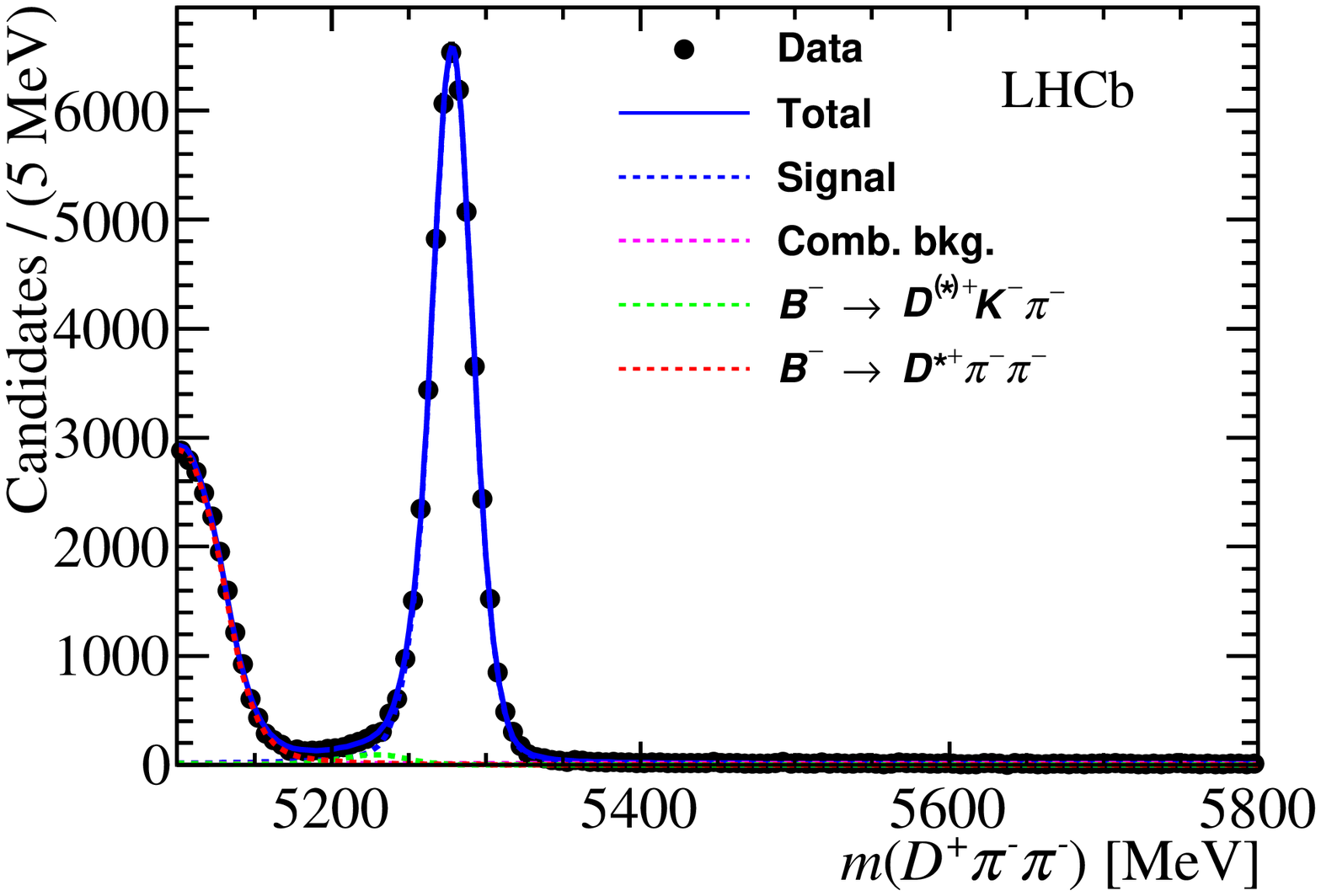}
  \caption{Left: $D^+ K^- \pi^-$ mass spectrum. Right: $B^- \to D^+ \pi^- \pi^-$ mass spectrum.}
   \label{fig:fig4}  
\end{figure}

\begin{figure}[htb]
      \centering
    \begin{overpic}[clip,width=1.5in]{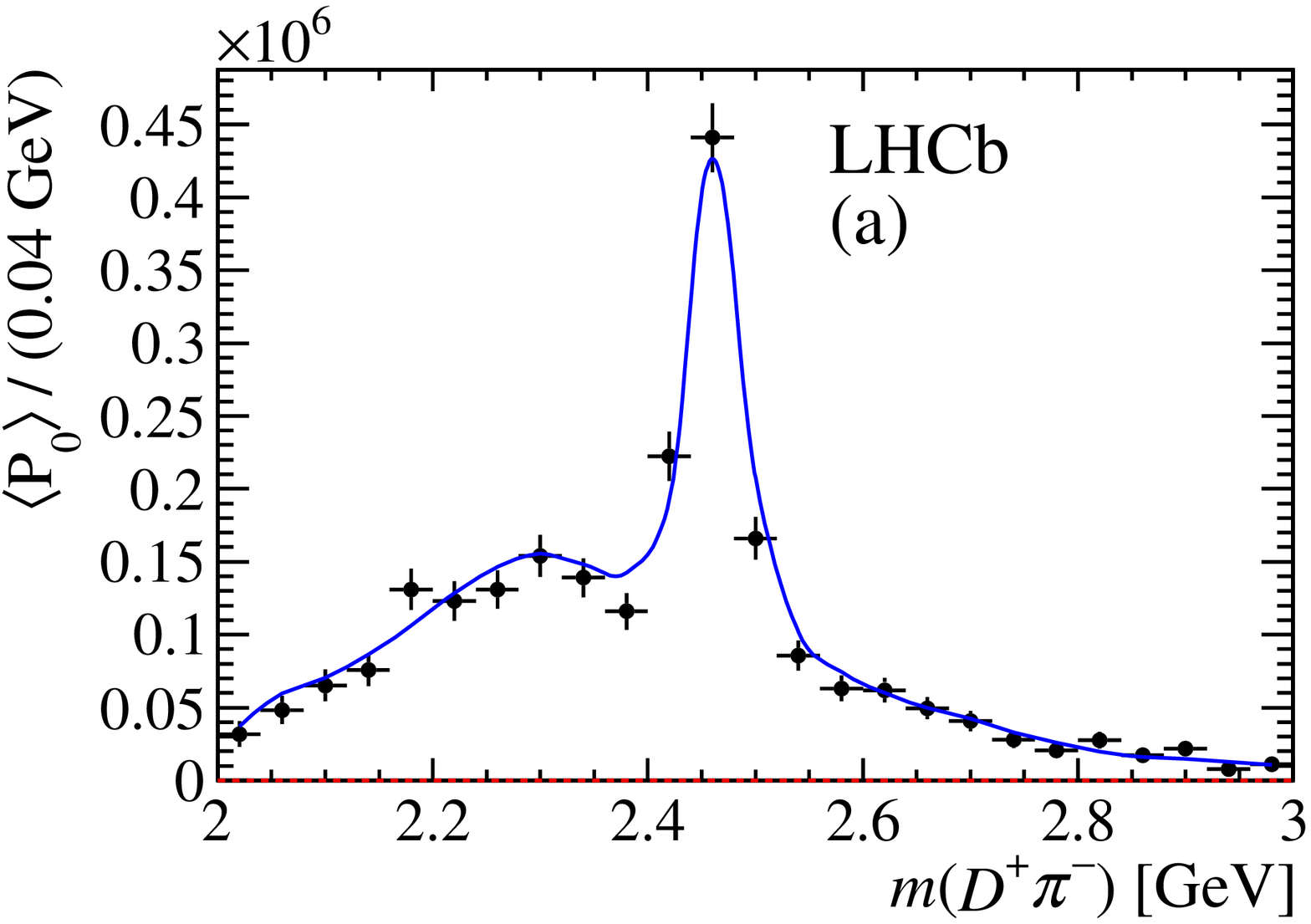}
      \put(70,40){\color{red}$P_0$}
    \end{overpic}
    \begin{overpic}[clip,width=1.5in]{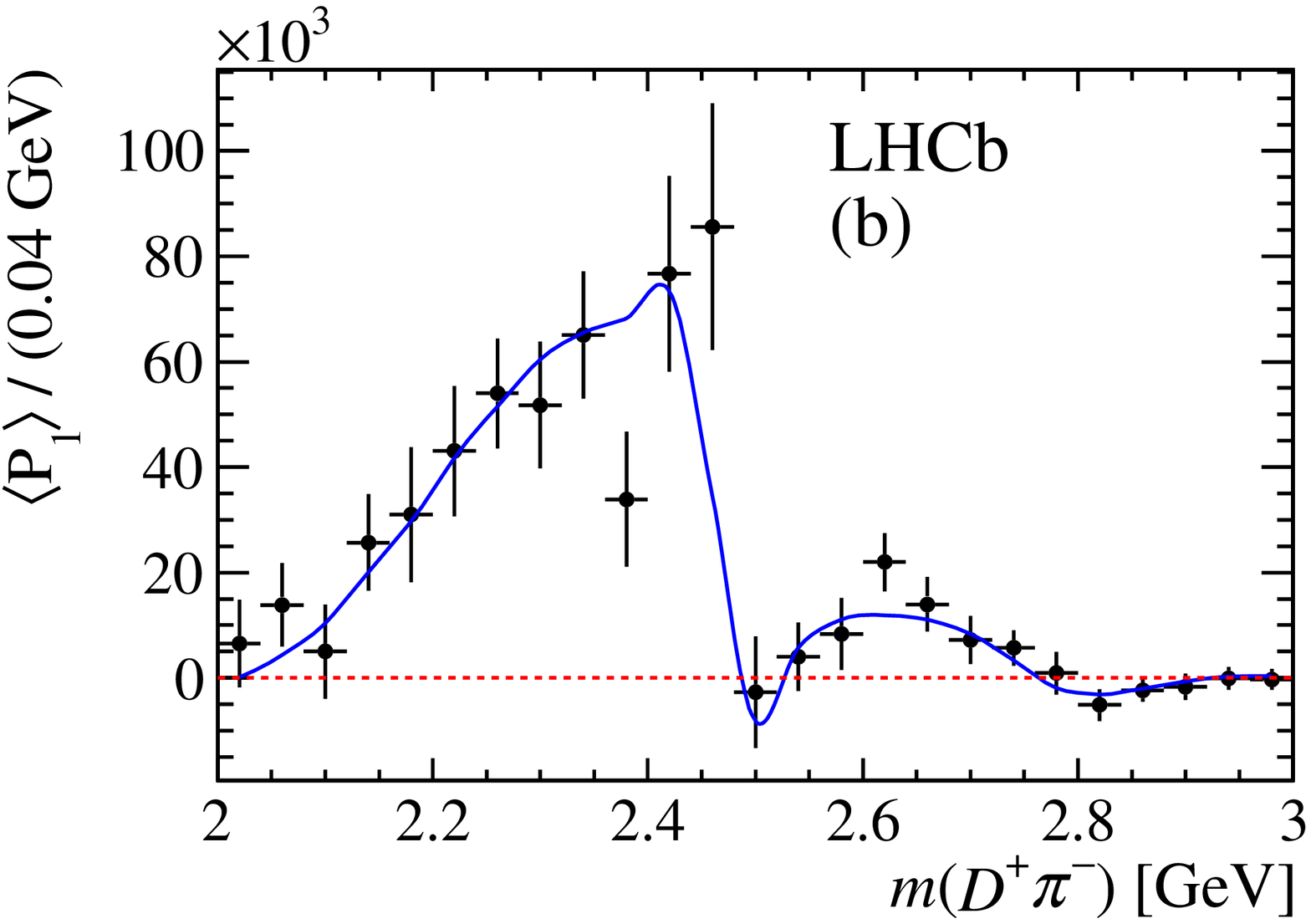}
      \put(70,40){\color{red}$P_1$}
    \end{overpic}    
        \begin{overpic}[clip,width=1.5in]{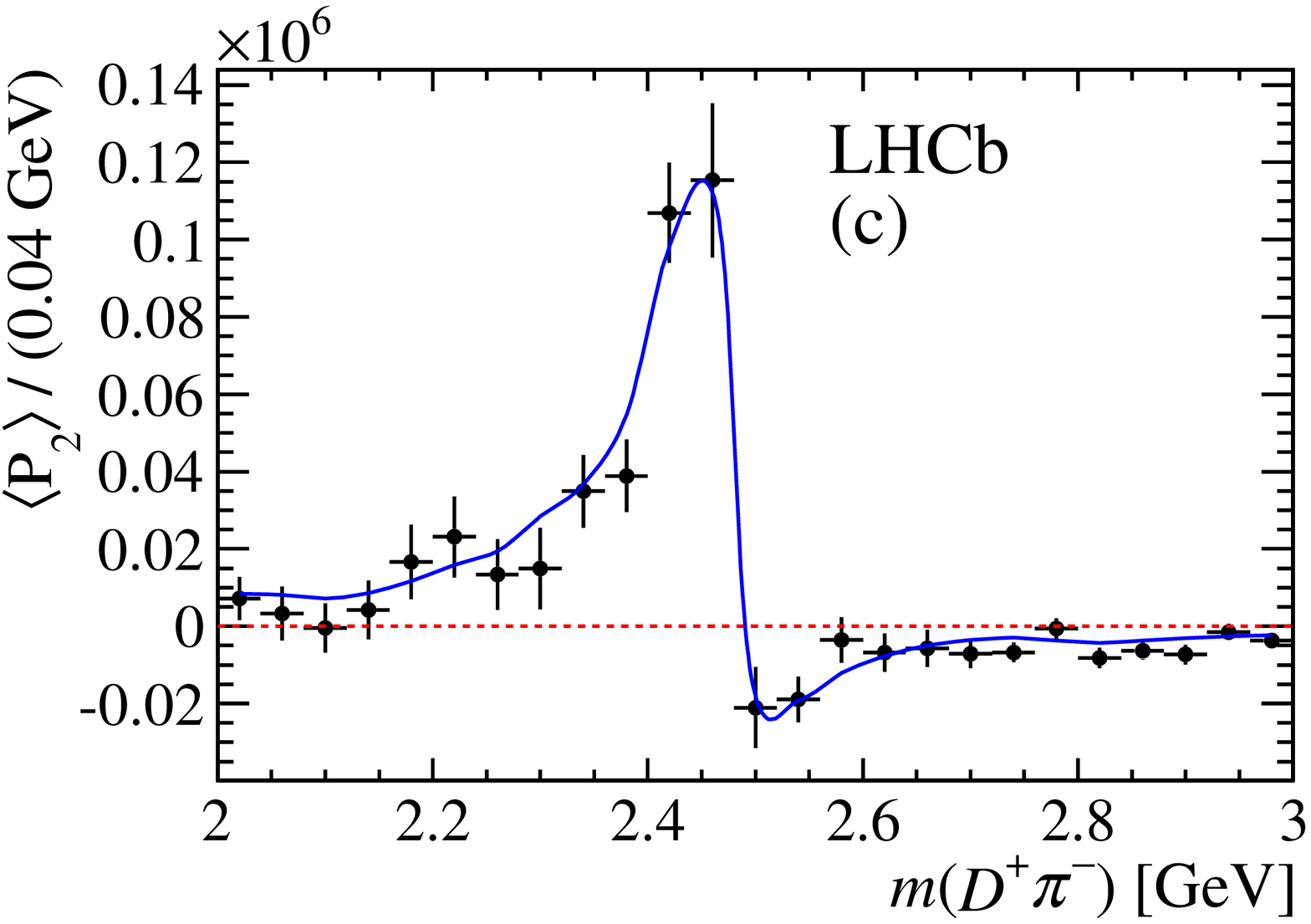}
      \put(70,40){\color{red}$P_2$}
        \end{overpic}
            \begin{overpic}[clip,width=1.5in]{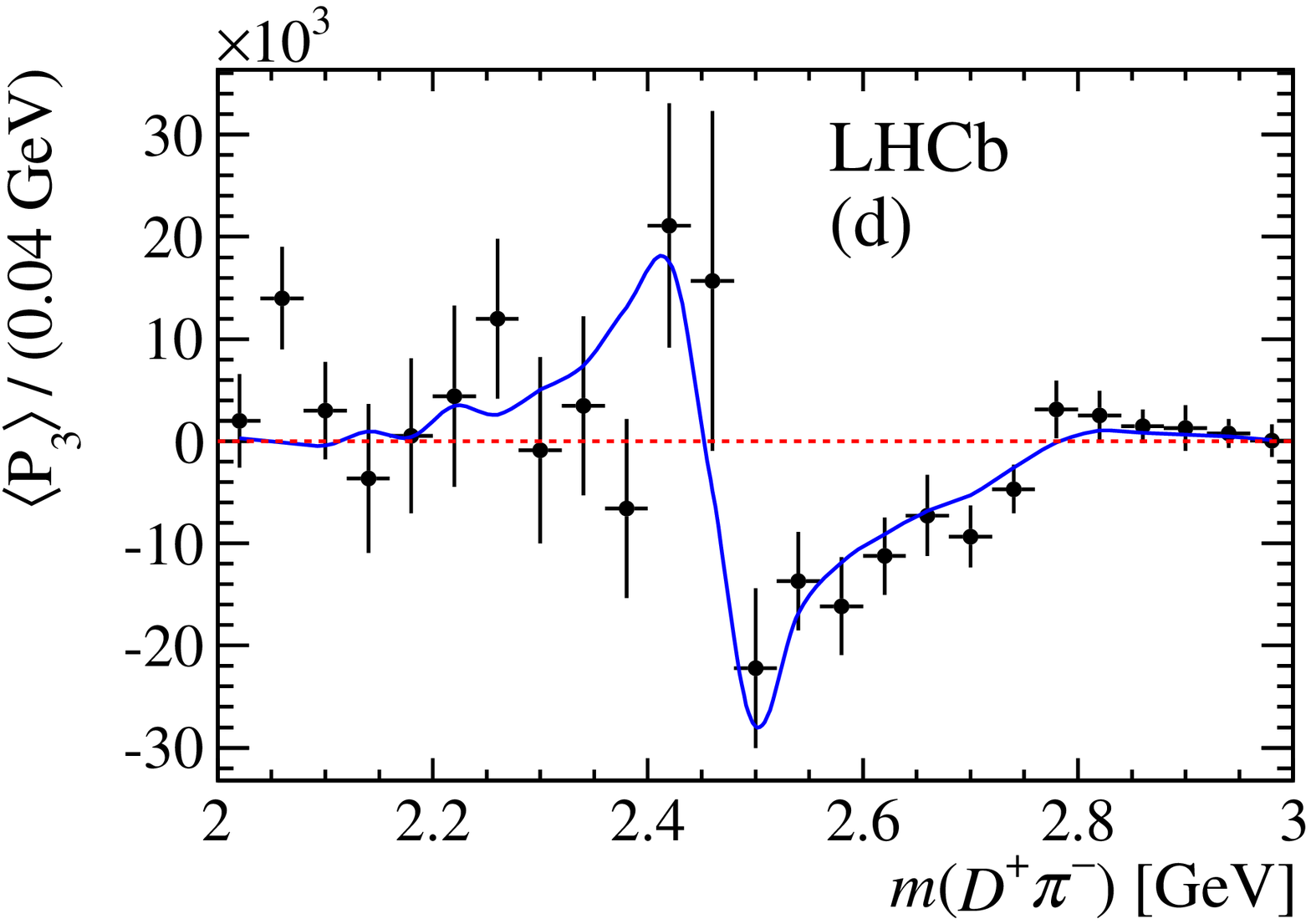}
      \put(70,15){\color{red}$P_3$}
            \end{overpic}
            \begin{overpic}[clip,width=1.5in]{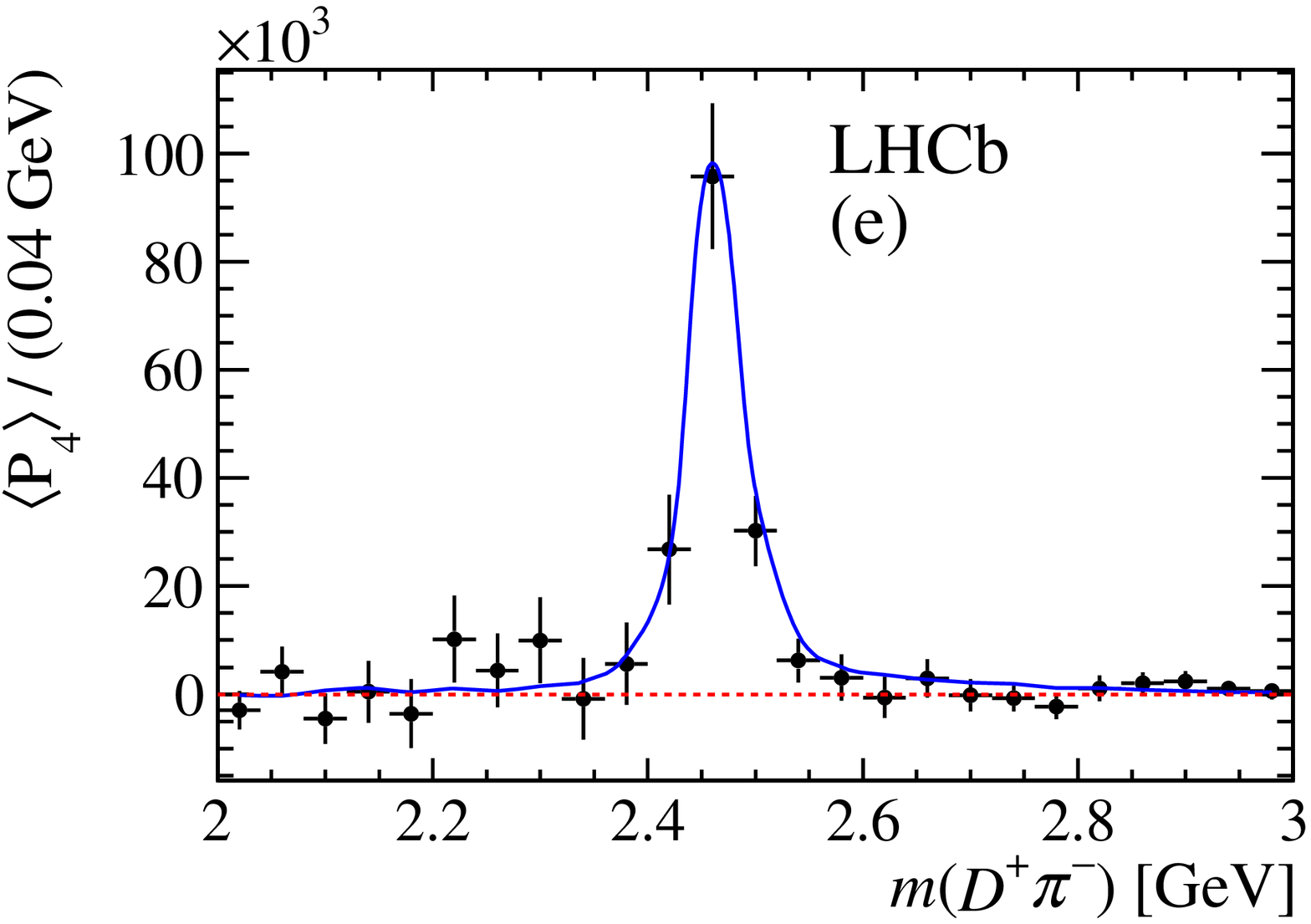}
      \put(70,40){\color{red}$P_4$}
            \end{overpic}
  \caption{$B^- \to D^+ K^- \pi^-$. $D^+ \pi^-$ mass spectrum, weighted by Legendre polynomial moments.}
   \label{fig:fig5}            
    \end{figure}

\begin{figure}[htb]
  \centering
    \includegraphics[width=3.0in]{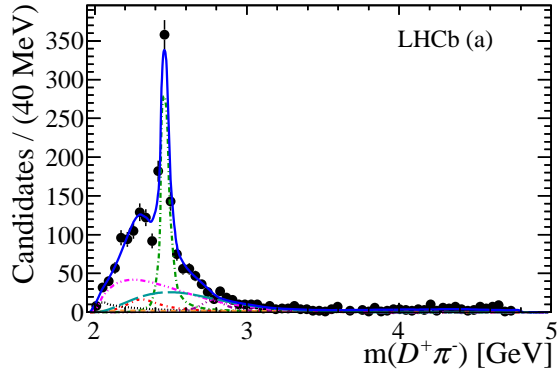}
  \caption{$B^- \to D^+ K^- \pi^-$. $D^+ \pim$ fit projection with fit result.}
   \label{fig:fig6}  
\end{figure}

      The $D^+ K^- \pi^-$ mass spectrum~\cite{dkpi} is shown in fig.~\ref{fig:fig4} (Left) and contains $\approx 2K$ events in the $B^-$ signal region.
To obtain a high $B^-$ signal purity we make use of neural network's trained by control samples, especially the $B^- \to D^+ \pi^- \pi^-$ final state which contains $\approx 49K$ events and shown in fig.~\ref{fig:fig4} (Right).
We perform a standard Dalitz plot analysis of the $B^- \to D^+ K^- \pi^-$ system.
In this final state, resonance production can only occur in the $D^+ \pi^-$ system. We plot, in fig.~\ref{fig:fig5}, the efficiency corrected and background subtracted $D^+ \pi^-$ mass spectrum, weighted by Legendre polynomial moments and compare with Dalitz plot fit results. 
    We note that $P_1$ is related to the S-P interference, while $P_3$ is related to the P-D interference.
We also observe, as expected, a Clear D-wave in $P_4$ due to the $D^{*}_{2}(2460)^0$. 
In the Dalitz analysis a better fit is obtained introducing virtual $D^{*}_{v}(2007)^{0}$, $B^{*0}_{v}$ and nonresonant contributions. We also observe a clear spin-2 $D^*_2(2460)^0$ signal and a $D^{*}_{J}(2760)^{0}$ spin-1 resonance.
The resulting resonance parameters and fitted fractions are given in Table~\ref{tab:tab1}.
\begin{table}
  \begin{center}
\begin{tabular}{l|cccc}
\hline
Resonance & Spin & Parameters & Fit fraction\\
\hline \\ [-3.0ex] 
$D^{*}_{0}(2400)^{0}$ &0& PDG & $\phantom{2}8.3 \pm 2.6 \pm 0.6 \pm \phantom{2}1.9$ \\
$D^{*}_{2}(2460)^{0}$ &2& $m=2464.0 \pm 1.4$, $\Gamma= 43.8 \pm 2.9\mev $& $31.8 \pm 1.5 \pm 0.9 \pm \phantom{2}1.4$\\ 
$D^{*}_{J}(2760)^{0}$ &1& $m =2781 \pm 18$, $\Gamma=  177 \pm 32\mev $& $\phantom{2}4.9 \pm 1.2 \pm 0.3 \pm \phantom{2}0.9$\\ 
\hline
\end{tabular}
\caption{$B^- \to D^+ K^- \pi^-$. Resonances parameters from the Dalitz analysis.}
\label{tab:tab1}
\end{center}
\end{table}
The $D^+ \pim$ fit projection with fit result is shown in Fig.~\ref{fig:fig6}.
No evidence for a $D^*_3(2760)$ spin-3 resonance is found in this final state.

\subsection{Dalitz plot analysis of $B^0 \to \bar D^0 \pi^+ \pi^-$}
    
\begin{figure}[htb]
        \centering
        \includegraphics[width=2.5in]{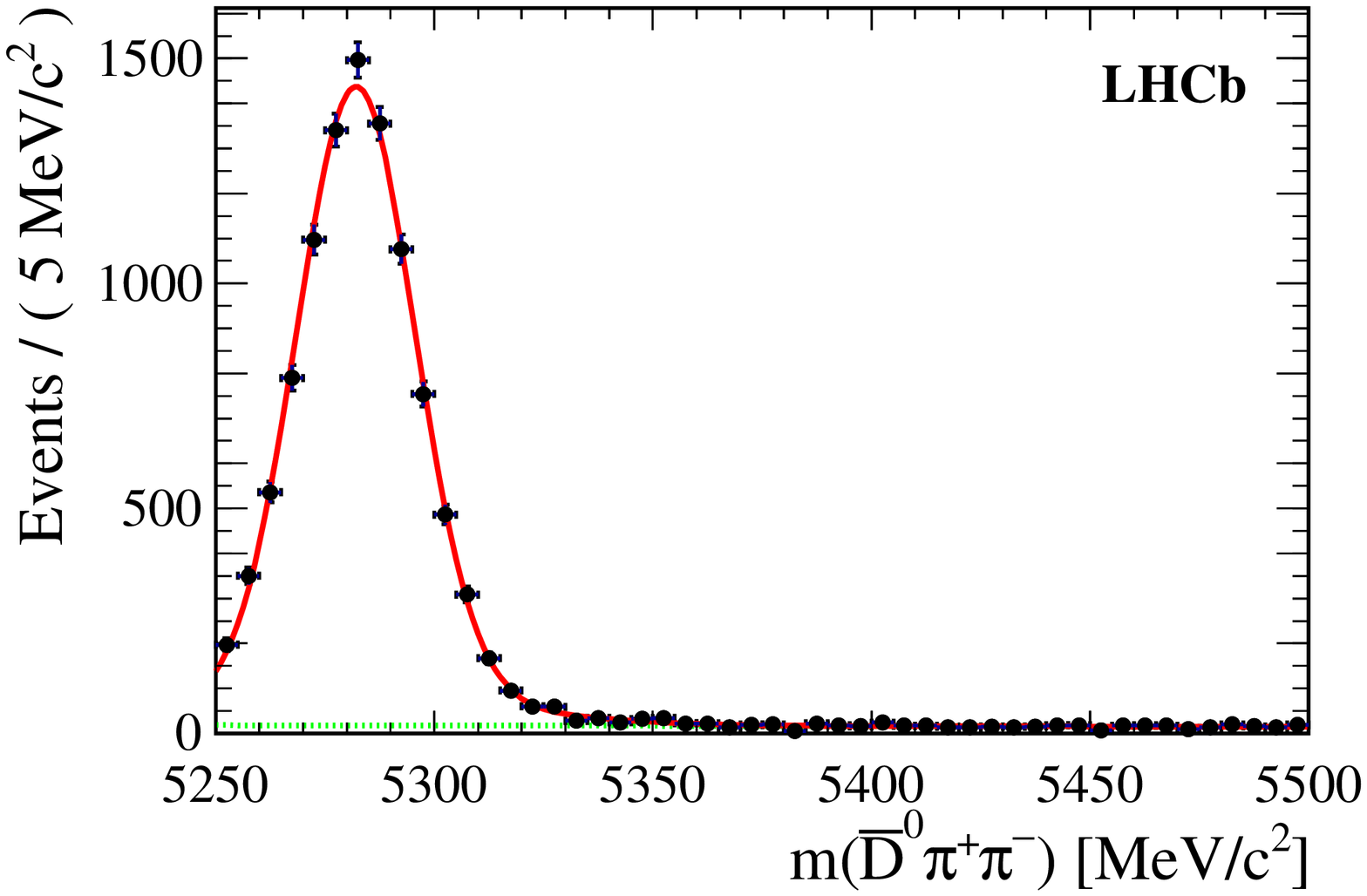}
        \includegraphics[width=2.5in]{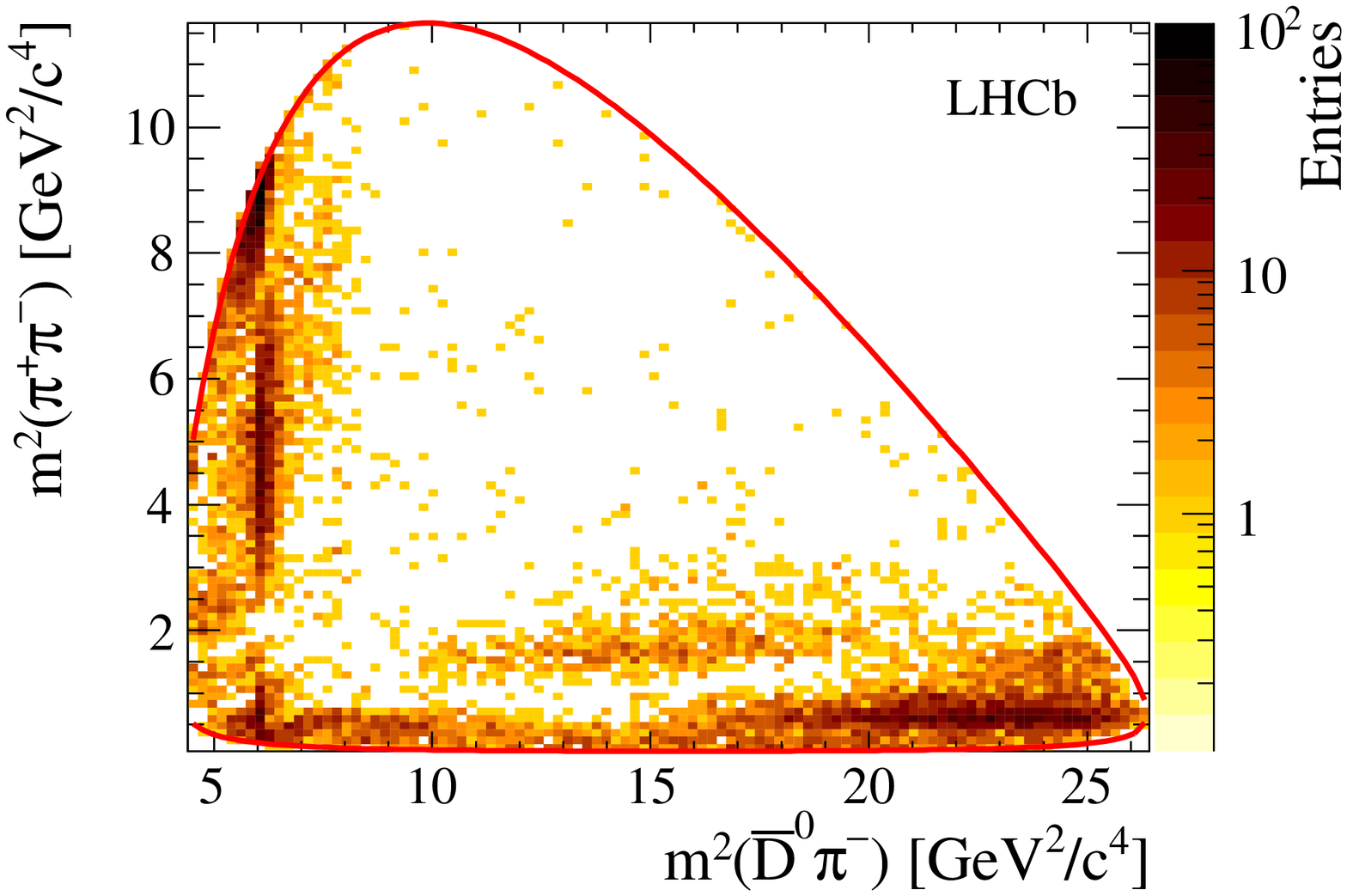}
  \caption{Left: $\bar D^0 \pi^+ \pi^-$ mass  spectrum. The green line represents the background contribution. Right:  $B^0 \to \bar D^0 \pi^+ \pi^-$ Dalitz plot.}
   \label{fig:fig7}         
      \end{figure}

\begin{figure}[htb]
\centering
  \includegraphics[width=2.5in]{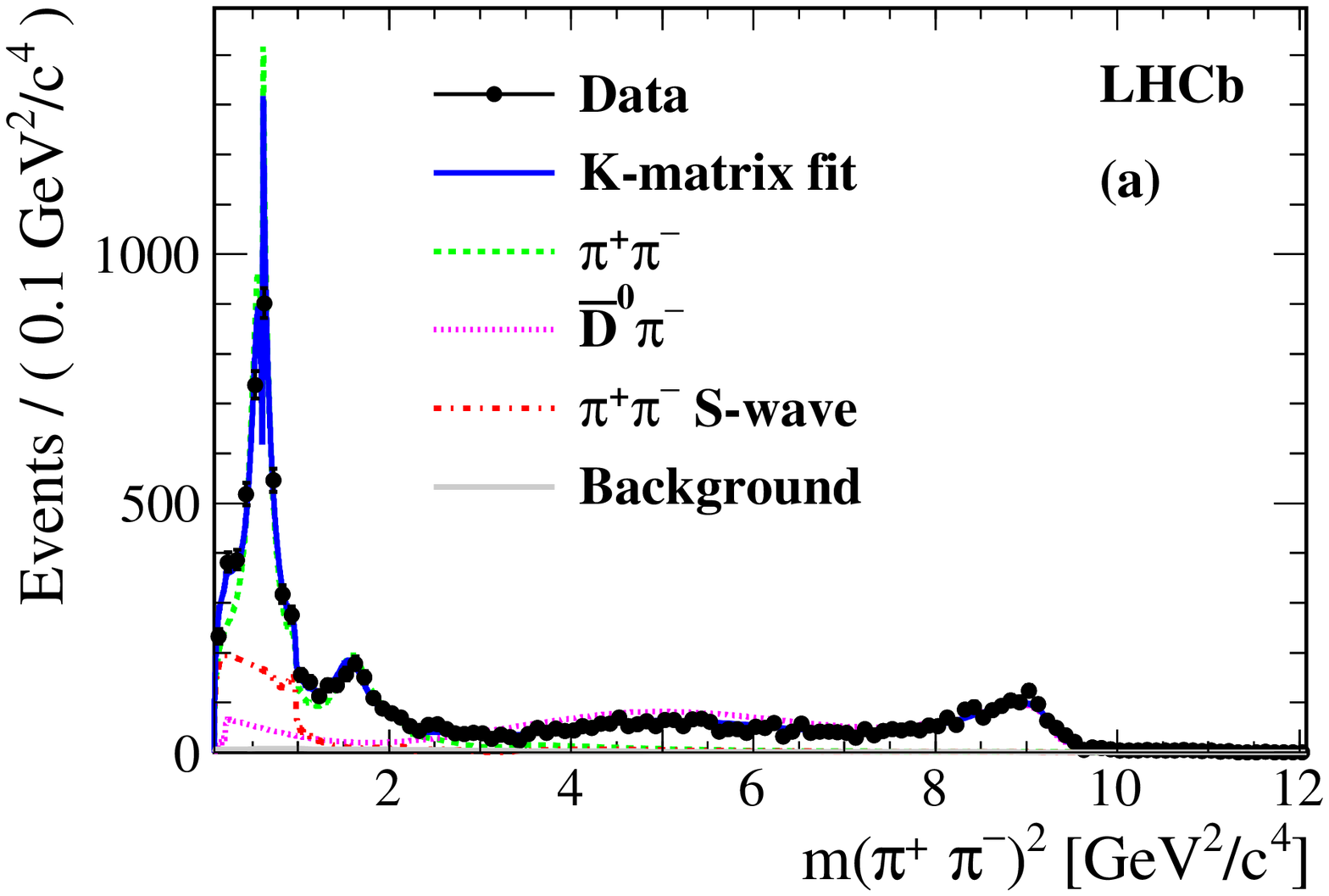}
  \includegraphics[width=2.5in]{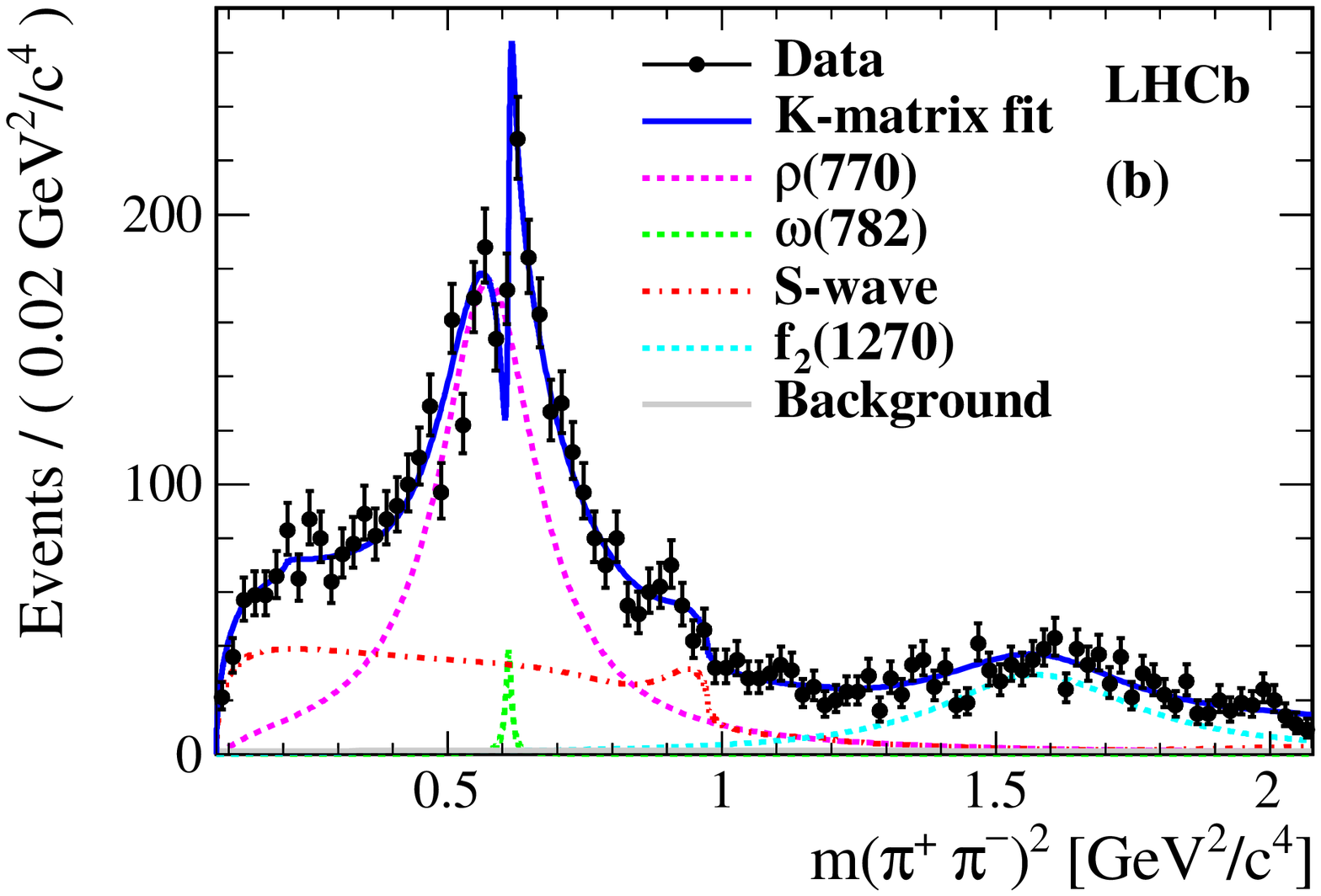}
  \caption{$B^0 \to \bar D^0 \pi^+ \pi^-$. $m^2(\pip \pim)$ fit projection with two different binnings.}
   \label{fig:fig8}   
      \end{figure}

\begin{figure}
\centering  
  \includegraphics[width=2.5in]{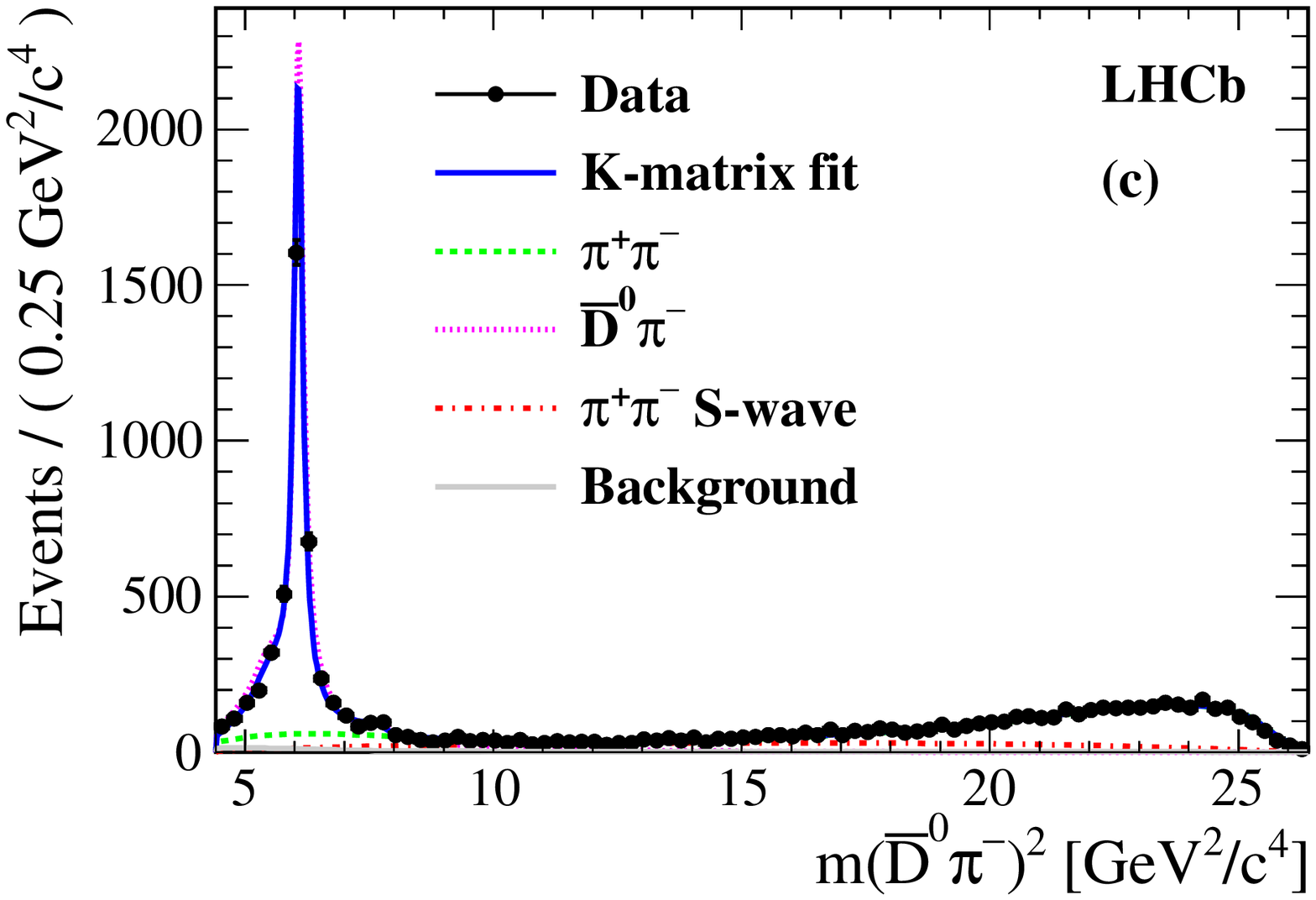}
  \includegraphics[width=2.5in]{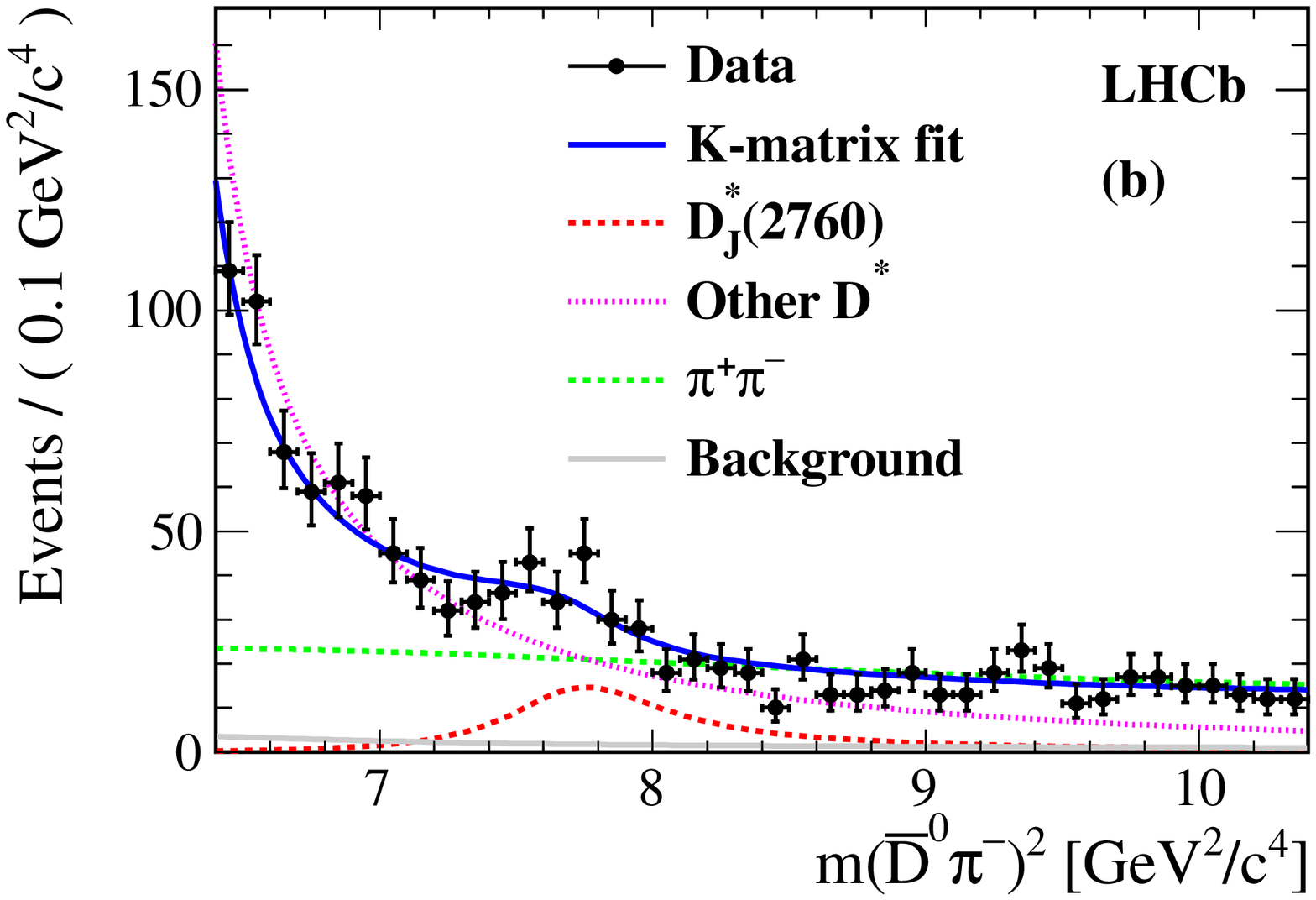}
  \caption{$B^0 \to \bar D^0 \pi^+ \pi^-$. $m^2(\bar D^0 \pi^-)$ fit projection with two different binnings.}
   \label{fig:fig9}   
\end{figure}

The $\bar D^0 \pi^+ \pi^-$ mass spectrum is shown in Fig.~\ref{fig:fig7} (Left)~\cite{dpipi}. The $B^0$ signal contains  9565 events with 97.8\% purity. The $B^0 \to \bar D^0 \pi^+ \pi^-$ Dalitz plot is shown in Fig.~\ref{fig:fig7} (Right).
    We observe the spin-2 $D^*_2(2460)^-$ signal along the $\bar D^0 \pim$ axis and the spin-1 $\rho(770)$ signal
along the $\pip \pim$ axis.
The Dalitz plot analysis has been performed using the isobar model and a K-matrix description of the $\pip \pim$ S-wave. Both methods give a good description of the data.
The $m^2(\pip \pim)$ fit projection are shown in Fig.~\ref{fig:fig8}. We observe a signal of $\rho/\omega$ interference.
The $m^2(\bar D^0 \pi^-)$ fit projection is shown in Fig.~\ref{fig:fig9}.      
The decay is dominated, in the $\pip \pim$ system, by S-wave ($16.51 \pm 0.70 \pm 1.68 \pm 1.10$)\% and $\rho(770)$ ($36.15 \pm 1.00 \pm 2.13 \pm 0.79$)\%. In the $\Dzb\pi^-$ system, the largest contribution comes from the $D_2^*(2460)^-$ resonance ($28.13 \pm 0.72 \pm 1.06 \pm 0.54$)\%.

The Dalitz plot analysis requires the presence of an additional $J^P=3^-$ resonance with a K-matrix model fitted fraction of
($1.58 \pm  0.22 \pm 0.18 \pm 0.07$)\%. The fitted resonances parameters 
are given in Table~\ref{tab:tab2}.
\begin{table}
    \begin{center}
\begin{tabular}{l| lc c}
\hline
 & &  Isobar & K-matrix \\
\hline
$D_0^*(2400)$ & $m$ &  $\al\all2349  \pm \al\all6 \pm \al\all1 \pm \al\all4$ & $\al\all2354 \pm \al\all7 \pm \all11 \pm \al\all2$   \\
& $\Gamma$ &  $\al\al\all217 \pm \all13 \pm \al\all5 \pm \all12$    &  $\al\al\all230 \pm \all15 \pm \all18 \pm \all11$   \\
$D_2^*(2460)$ & $ m$ & $2468.6 \pm 0.6 \pm 0.0 \pm 0.3$ & $2468.1 \pm 0.6 \pm 0.4 \pm 0.3$   \\
& $\Gamma$ &  $\al\al47.3 \pm 1.5 \pm 0.3 \pm 0.6$  &  $\al\al46.0 \pm 1.4 \pm 1.7 \pm 0.4$  \\
$D_3^*(2760)$ & $m$ &   $\al\all2798 \pm \al\all7 \pm \al\all1 \pm \al\all7$  &  $\al\all2802 \pm \all11 \pm \all10 \pm \al\all3$ \\
& $\Gamma$   & $\al\al\all105 \pm \all18 \pm \al\all6 \pm \all23$ & $\al\al\all154 \pm \all27 \pm \all13 \pm \al\all9$  \\
\hline
\end{tabular}
\caption{$B^0 \to \bar D^0 \pi^+ \pi^-$. Resonances parameters from the Dalitz analysis.}
\label{tab:tab2}
\end{center}
\end{table}

Comparing the $D_J^*(2760)$ parameters between inclusive and $B$ decays production, we observe some disagreement.
Both  $J^P=1^-$ and  $J^P=3^-$ resonances are expected in this mass region and inclusive data cannot separate
the two natural parity contributions.

\subsection{Dalitz plot analysis of $B^0 \to \bar D^0 K^+ \pi^-$}

\begin{figure}[htb]
        \includegraphics[width=3.0in]{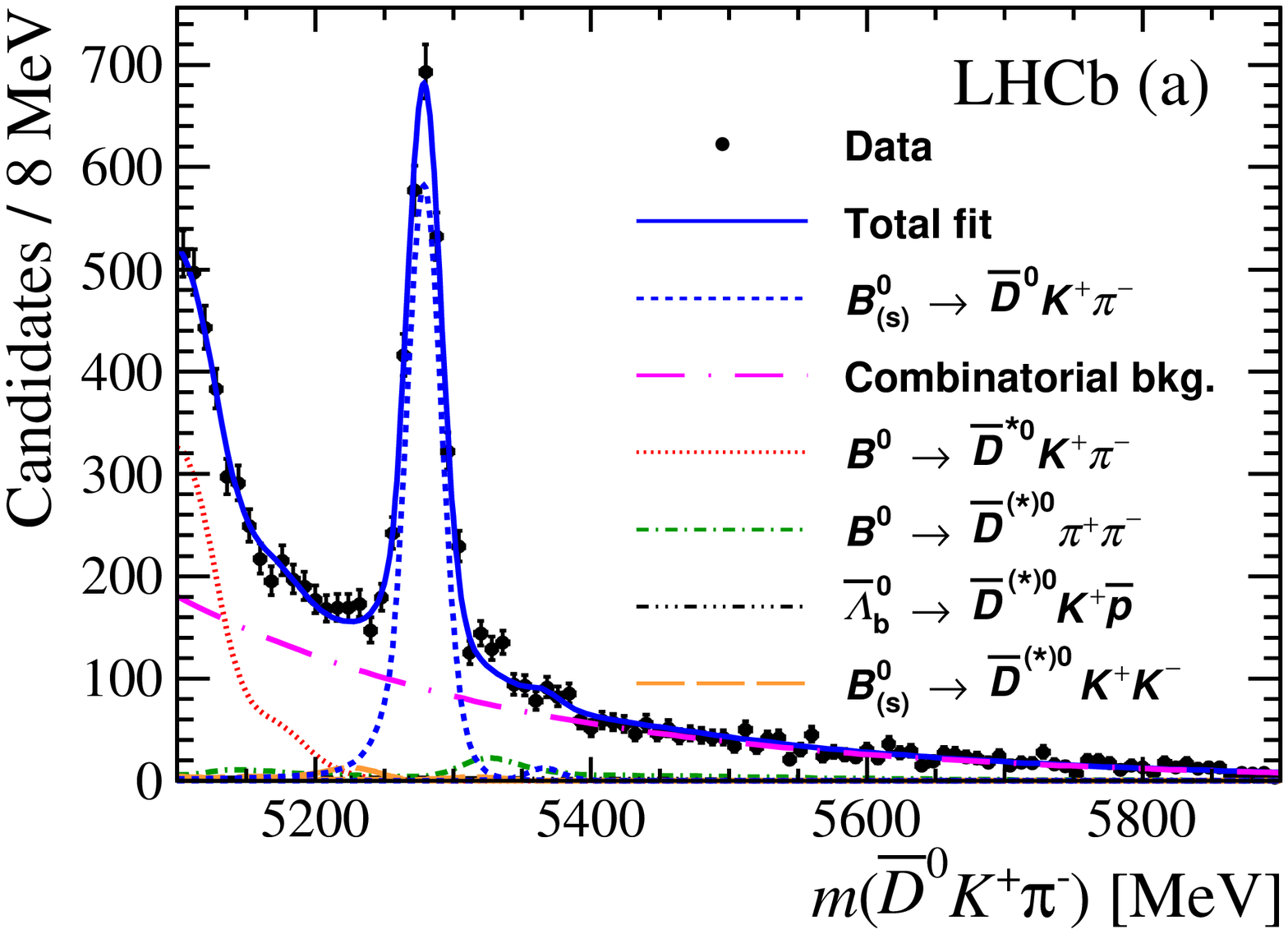}
        \includegraphics[width=3.0in]{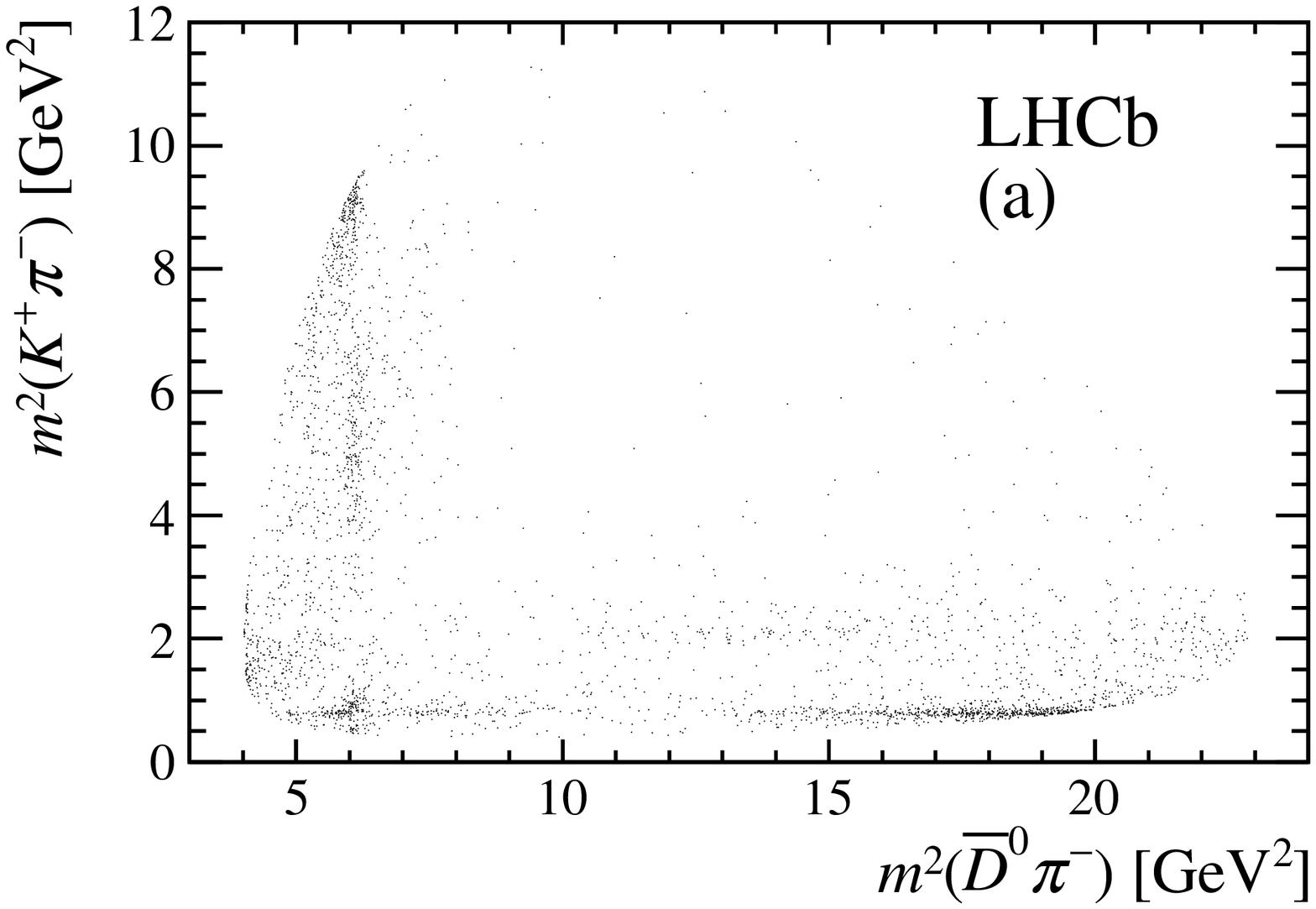}
  \caption{Left: $\bar D^0 K^+ \pi^-$  mass  spectrum. Right: $B^0 \to \bar D^0 K^+ \pi^-$ Dalitz plot.}
   \label{fig:fig10}           
      \end{figure}

\begin{figure}[htb]
        \centering
  \includegraphics[width=2.5in]{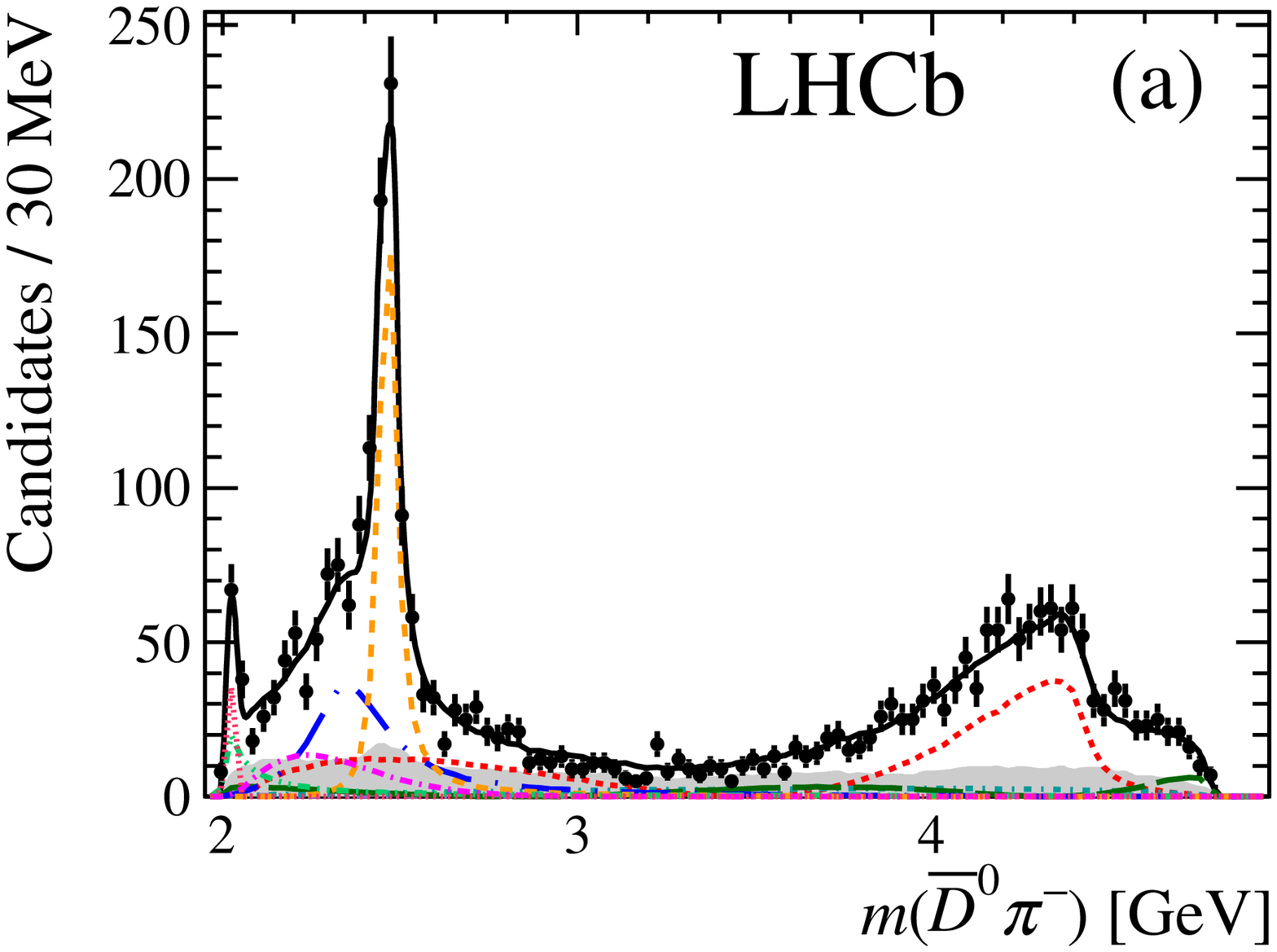}
  \includegraphics[width=2.5in]{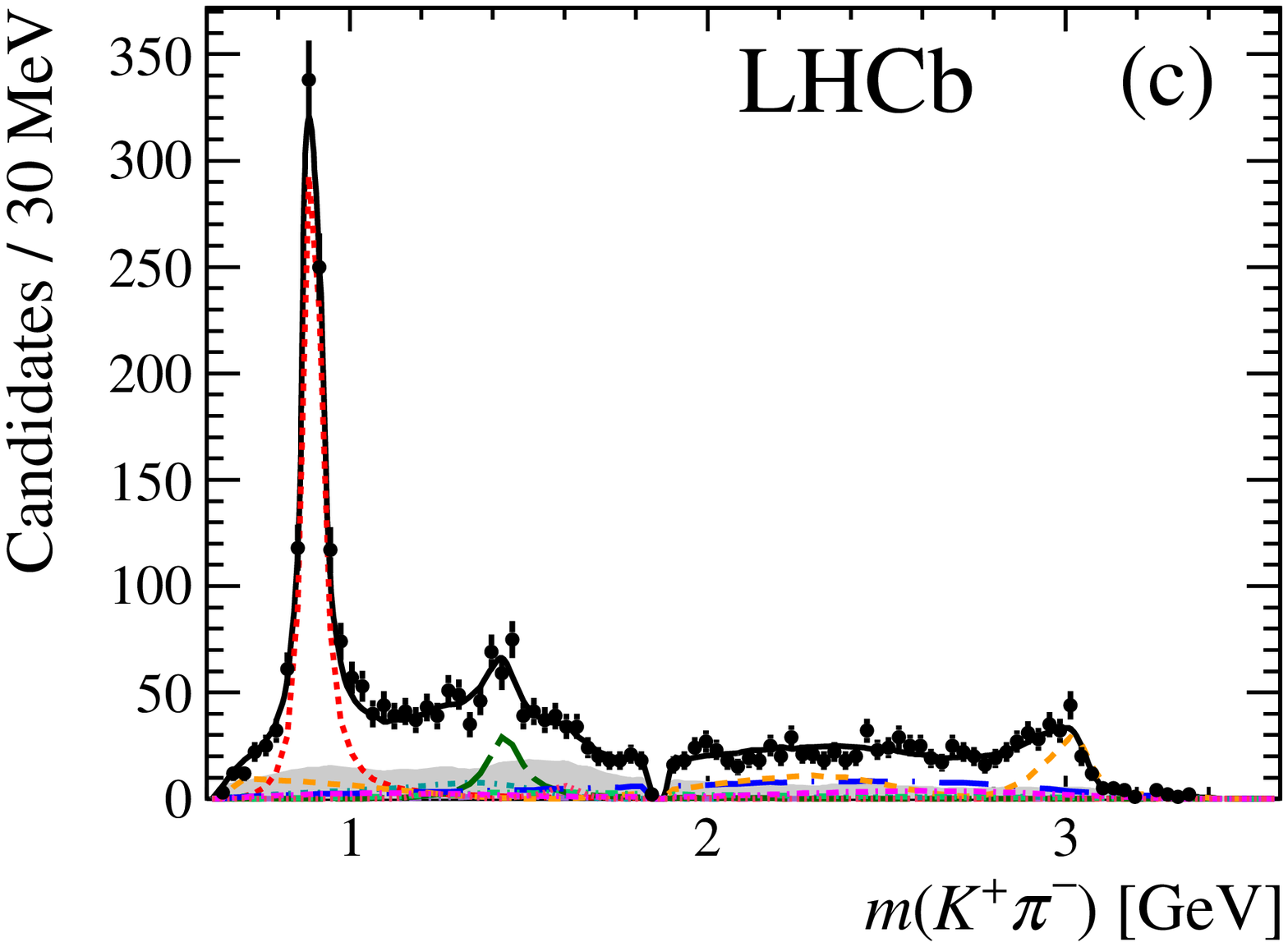}
    \caption{$B^0 \to \bar D^0 K^+ \pi^-$. $m(\bar D^0 \pi^-)$ and $m(K^+ \pi^-)$ fit projections.}
   \label{fig:fig11}
      \end{figure}

The $\bar D^0 K^+ \pi^-$ mass spectrum is shown in Fig.~\ref{fig:fig10}(Left)~\cite{dkpi0}. The $B^0$ signal region contains 2344 events. The fit projections are shown in Fig.~\ref{fig:fig11}.
The decay is dominated by intermediate resonance production of $\Kstar(892)^{0}$ ($37.4 \pm 1.5$)\%, $D^{*}_{0}(2400)^{-}$ ($19.3 \pm 2.8$)\% and $D^{*}_{2}(2460)^{-}$ ($23.1 \pm 1.2$)\%.       
These Dalitz analyses obtain new parameters for the broad $D^{*}_{0}(2400)^-$ resonance which are summarized in Table~\ref{tab:tab3}.
\begin{table}
  \begin{center}
\begin{tabular}{l|ccc}
\hline
Final state & Method & Mass & Width \\
\hline \\ [-2.5ex]
$B^0 \to \bar D^0 K^+ \pi^-$ & Free & $ 2360 \pm 15 $ & $ 255 \pm 26 $\\
$B^0 \to \bar D^0 \pi^+ \pi^-$ & Free &  $2354 \pm 7 $ & $230 \pm 15 $ \\
$B^- \to D^+ K^- \pi^-$& (PDG) & $2318 \pm 29 $ & $267 \pm 40$ \\
\hline
\end{tabular}
\caption{$D^{*}_{0}(2400)$ resonances parameters from the different Dalitz analyses.}
\label{tab:tab3}
\end{center}
\end{table}
No evidence is found for additional spin-1 or spin-3 $D^*_J$ resonances.

\section{Results on $D_{sJ}$ mesons spectroscopy}

\subsection{Inclusive studies}

\begin{figure}[htb]
  \centering
          \includegraphics[width=2.0in]{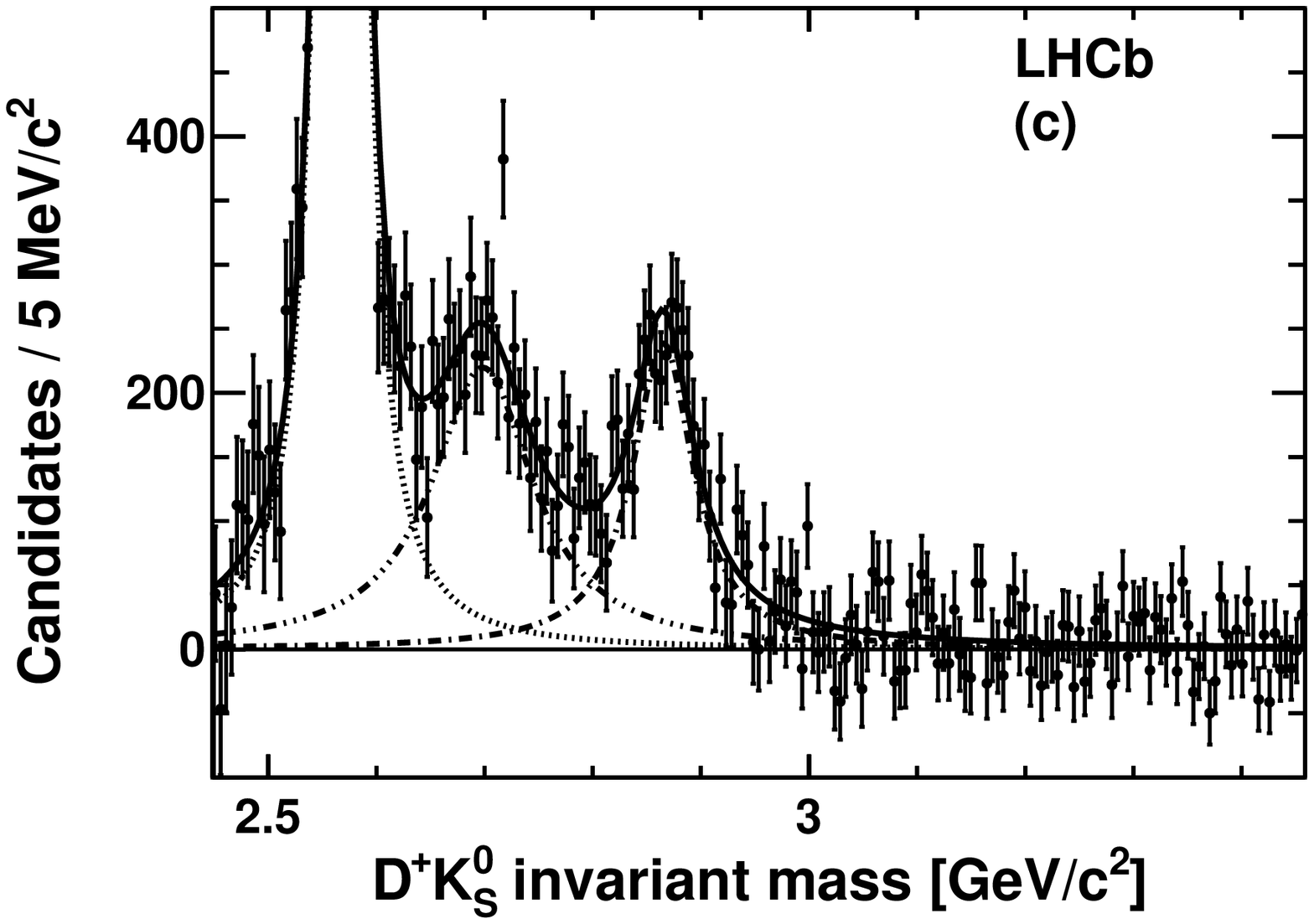}
          \includegraphics[width=2.0in]{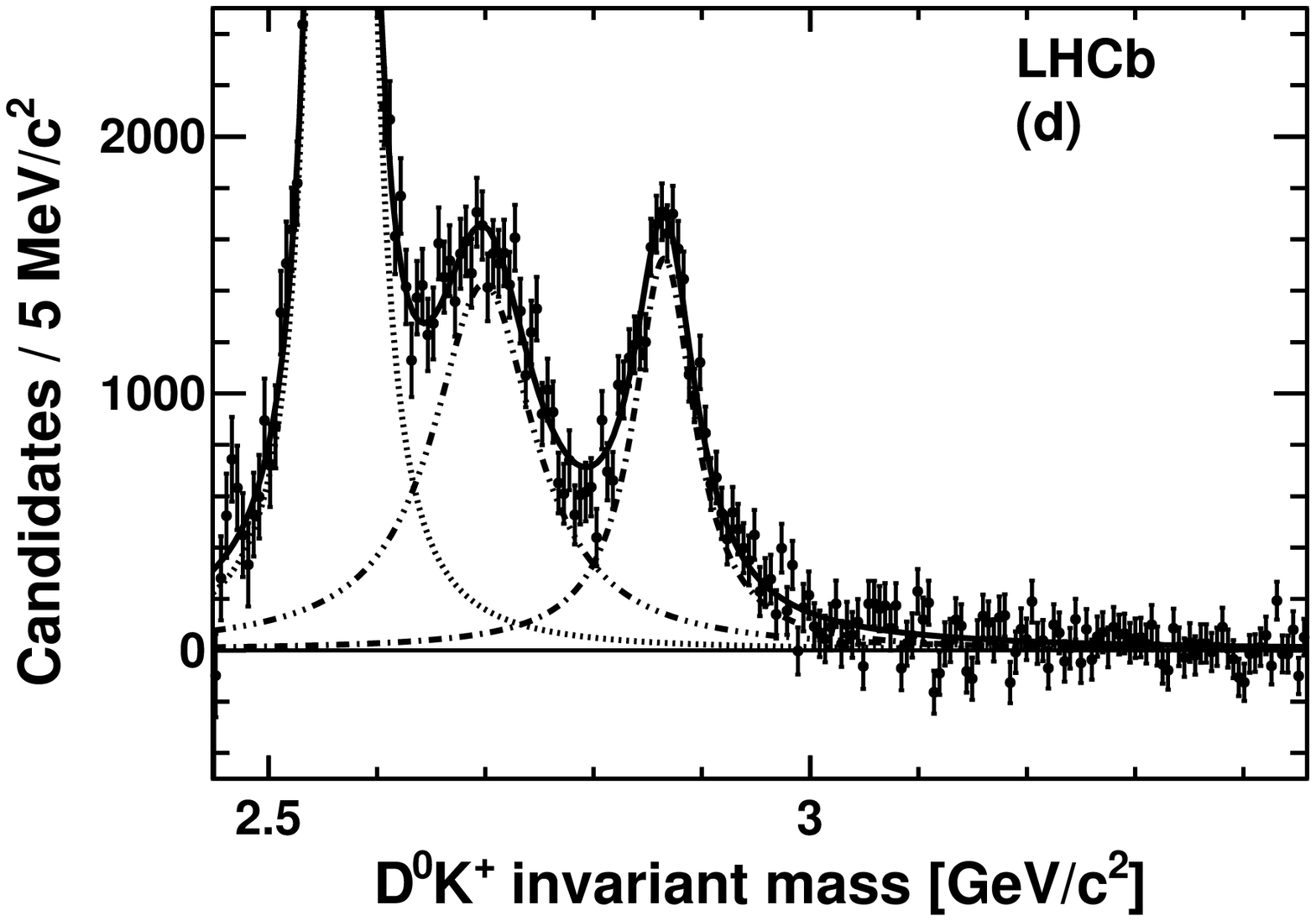}          
  \caption{Background subtracted (c) $D^+\KS$ and (d) $D^0 \Kp$ mass spectra.} 
   \label{fig:fig12}           
      \end{figure}

Using samples of $0.36\times10^6$ $D^+\KS$ and $3.15\times10^6$ $D^0K^+$ inclusive candidates, the $D_{s1}^*(2710)^+$ and $D_{sJ}^*(2860)^+$ have been observed with parameters consistent with previous measurements~\cite{dk}. The background subtracted $D^+\KS$ and $D^0 \Kp$ mass spectra are shown in Fig.~\ref{fig:fig12}.
\subsection{Dalitz plot analysis of $B_s^0 \to \Dzb\Km\pip$}

\begin{figure}[htb]
  \centering
  \includegraphics[width=2.0in]{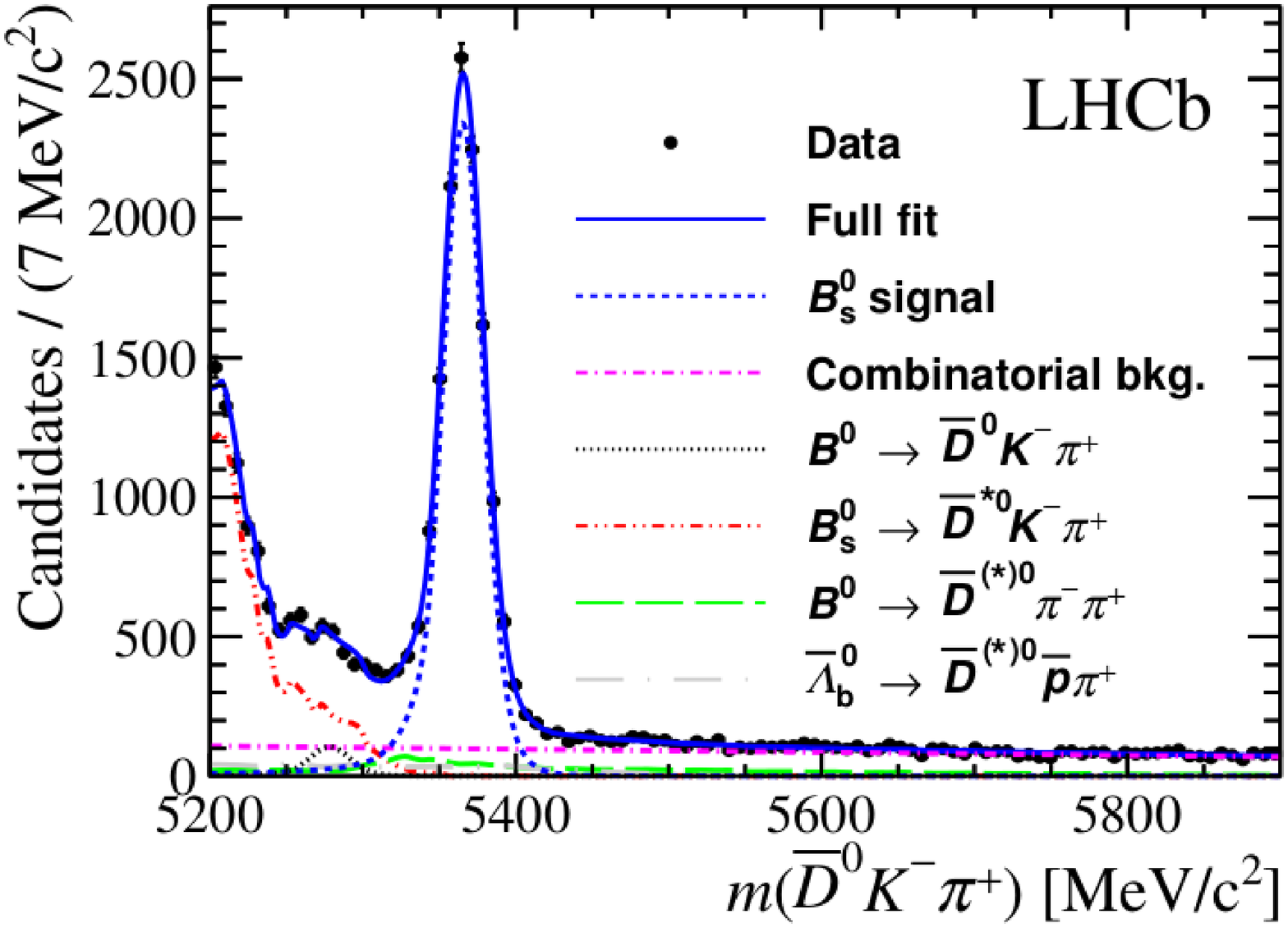}
  \includegraphics[width=3.7in]{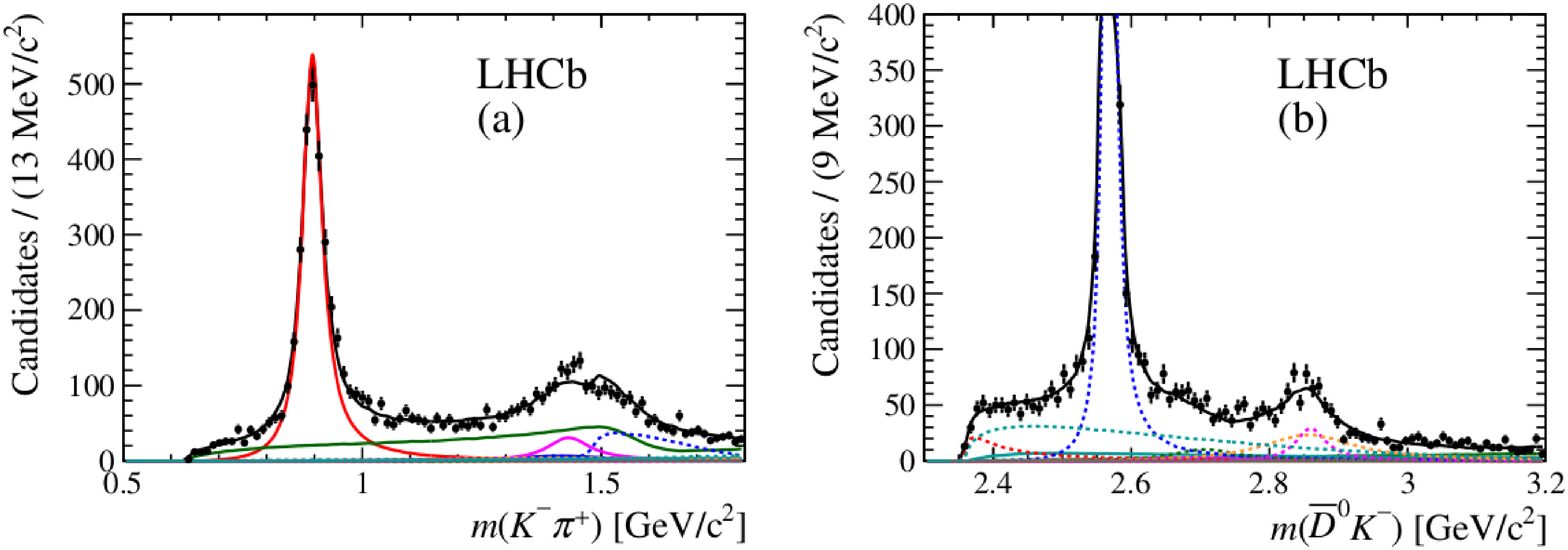}
  \caption{(Left) $\Dzb\Km\pip$ mass spectrum. (Right) $K^- \pip)$ and $\Dzb\Km$ mass projections.}
   \label{fig:fig13}
\end{figure}

Fig.~\ref{fig:fig13}(Left) shows the $\Dzb\Km\pip$ mass spectrum~\cite{bs}. The $B_s^0$ signal contains
   $11 302 \pm 159$ signal events. 
The Dalitz analysis shows that the largest components are:
$\Kstarb(892)^{0}$ ($28.6\,\%$),
$D^{*}_{s2}(2573)^{-}$ ($25.7\,\%$),
$K \pi$ S-wave (LASS) ($21.4\,\%$)
$\bar D^0 \Km$ nonresonant ($12.4\,\%$).
A signal present in the 2860 MeV $\Dzb \Km$ mass region which is 
described by a superposition of a spin-1 $(5.0 \pm 1.2 \pm 0.7 \pm 3.3)\,\%$
and a spin-3 $(2.2 \pm 0.1 \pm 0.3 \pm 0.4)\,\%$ resonance. The fitted resonances parameters are:
$  m(D^{*}_{s1}(2860)^-)   =    2859    \pm 12   \pm 6    \pm 23  \mevcc, \Gamma(D^{*}_{s1}(2860)^-) = 159      \pm 23   \pm 27   \pm 72  \mevcc $
$  m(D^{*}_{s3}(2860)^-)       =  2860.5   \pm 2.6  \pm 2.5  \pm 6.0 \mevcc,\Gamma(D^{*}_{s3}(2860)^-)  =  53       \pm 7    \pm 4    \pm 6   \mevcc $.

\subsection{Determination of the $X(3872) \to J/\psi \rho(770)$ quantum numbers}
\begin{figure}
       \centering
      \includegraphics[width=2.5in]{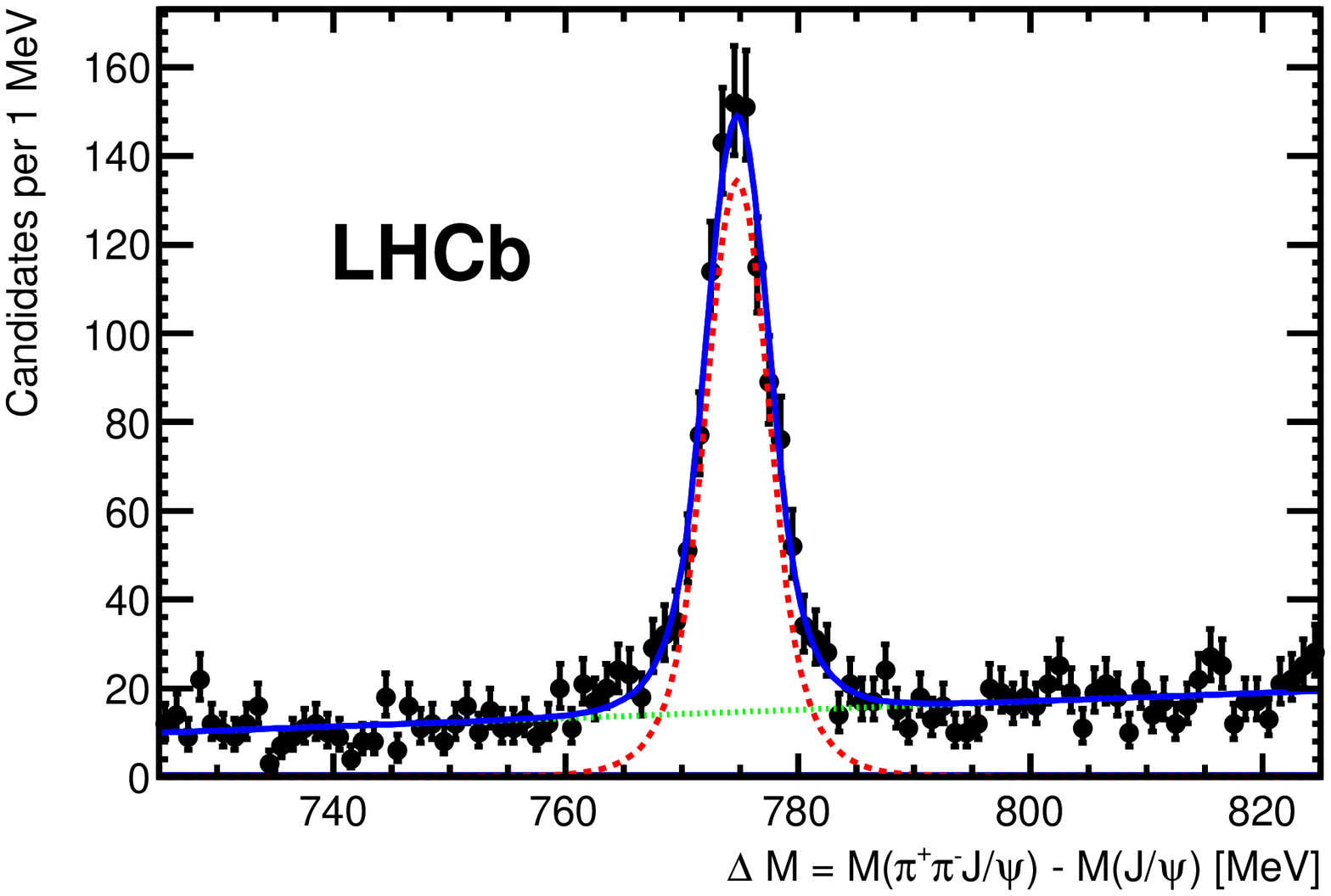}
      \includegraphics[width=3.0in]{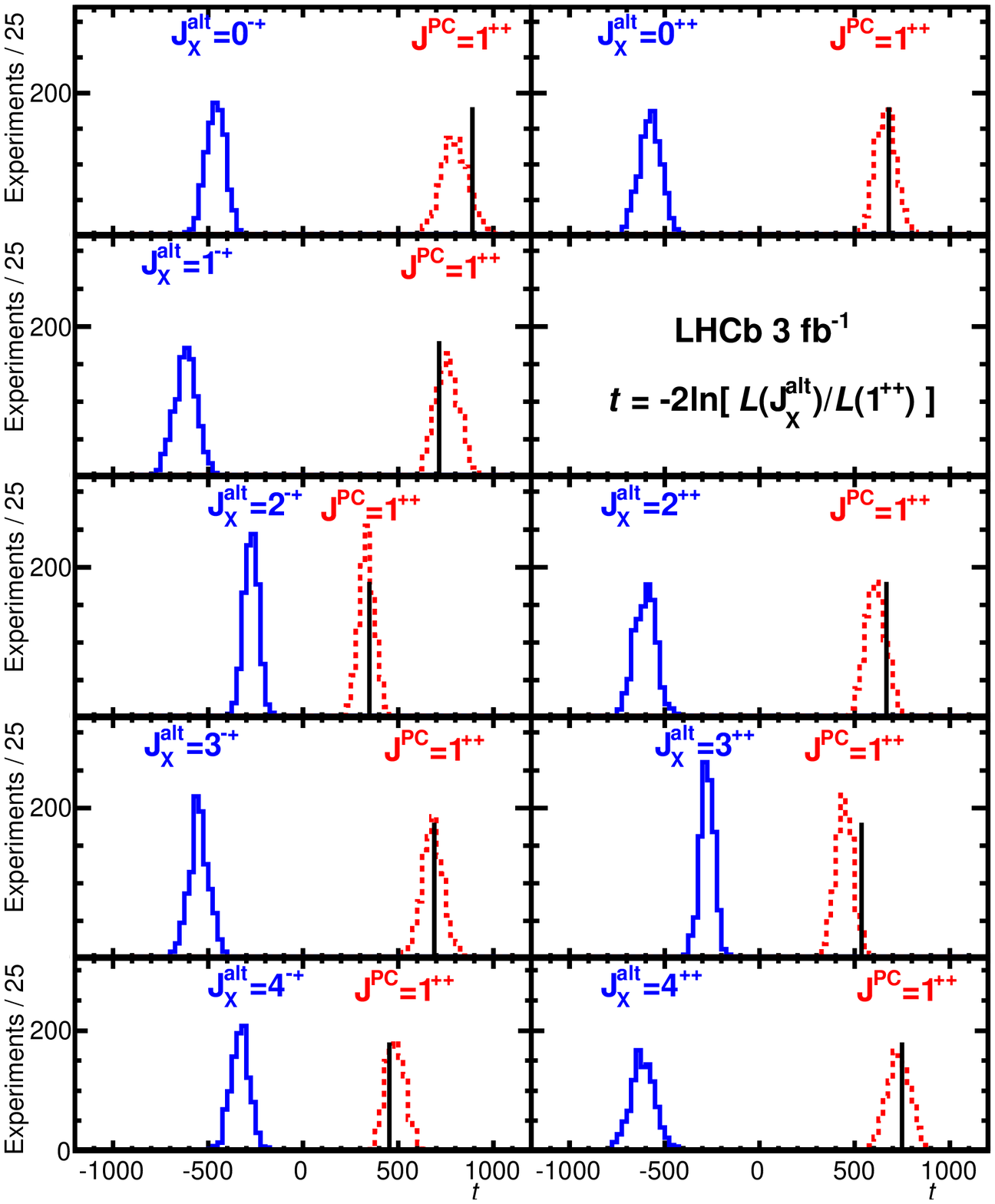}
  \caption{(Left) X(3872) signal. (Right) Distributions of the test statistic for different spin-parity hypothesis.}
   \label{fig:fig14}      
    \end{figure}

     We study the decay $B^+ \to X(3872) K^+$ with $X(3872) \to J/\psi \pip \pim$~\cite{x3872}~\cite{x3872}.
     The quantum numbers of $X(3872) \to J/\psi \rho(770)$  have been determined to be $J^{PC}=1^{++}$.
     However it was assumed that the decay is dominated by the lowest values of angular momentum between the $X(3872)$ decay products.
     The analysis is repeated here using 3-times the statistics and without any assumption on $L_{min}$. The $\Delta M$ signal for $ J/\psi \pip \pim$ is shown in Fig.~\ref{fig:fig14}. The $B^+$ signal for $B^+ \to X(3872) K^+$ contains $1011 \pm 38$ with 80\% signal purity.
    
     The distributions of the test statistic $t \equiv -2\ln[\L(J_X^{\rm alt}))/\L(1^{++})]$, for the simulated experiments under the $J^{PC}=J_X^{\rm alt}$ hypothesis and under the $J^{PC}=1^{++}$ hypothesis are
     shown in Fig.~\ref{fig:fig14}
The $J^{PC}=1^{++}$ hypothesis gives the highest Likelihood value with an upper limit of $D$-wave contribution of 4\% at 95\% C.L.

\end{document}

%% file: econfmacros.tex
\def\beq{\begin{equation}}
\def\eeq#1{\label{#1}\end{equation}}
\def\eeqn{\end{equation}}

\def\beqa{\begin{eqnarray}}
\def\eeqa#1{\label{#1}\end{eqnarray}}
\def\eeqan{\end{eqnarray}}

\let\bar=\overbar

\def\D{{\cal D}}
\def\L{{\cal L}}

\def\Dslash{\not{\hbox{\kern-4pt $D$}}}
\def\dslash{\not{\hbox{\kern-2pt $\del$}}}

\def\msb{{\bar{\ssstyle M \kern -1pt S}}}

